\documentclass[a4paper,11pt]{article}
\usepackage{jcappub} 
\usepackage{float}
\usepackage{cancel}
\usepackage{upgreek}
\usepackage{booktabs}
\usepackage{siunitx}
\usepackage{placeins}
\usepackage{orcidlink}
\usepackage{comment}
\usepackage[normalem]{ulem}
\usepackage{booktabs}
\usepackage{appendix}
\usepackage{indentfirst}
\usepackage{xcolor}
\usepackage{amsmath,mathtools}
\usepackage[T1]{fontenc} 
\usepackage{hyperref}
\usepackage{tikz}
\usepackage{booktabs}

\newcommand{\INFNNA}{INFN Sezione di Napoli, Università degli Studi di Napoli “Federico II”,\\Complesso Universitario di Monte Sant'Angelo, Edificio G, Via Cinthia, I-80126, Napoli, Italy}
\newcommand{\DIPFISNAP}{Dipartimento di Fisica “E. Pancini”, Università degli Studi di Napoli “Federico II”,\\Complesso Universitario di Monte Sant'Angelo, Edificio G, Via Cinthia, I-80126, Napoli, Italy}
\newcommand{\CCA}{Center for Computational Astrophysics, Flatiron Institute, 162 5th Avenue, 10010, New York, NY, USA
}
\newcommand{\ICTP}{ICTP, International Centre for Theoretical Physics, Strada Costiera 11, 34151, Trieste, Italy}

\author[a,b]{Tommaso Moretti,\orcidlink{0009-0006-4815-4764}}
\author[a,b]{Noemi Frusciante, \orcidlink{0000-0002-7375-1230}}
\author[c,d,1]{\\Giovanni Verza \orcidlink{0000-0002-1886-8348}\note{This paper is dedicated to the memory of our friend and colleague Giovanni who passed away on 26 June 2026 just before the submission of the present work.}}

\affiliation[a]{\DIPFISNAP}
\affiliation[b]{\INFNNA}
\affiliation[c]{\ICTP}
\affiliation[d]{\CCA}

\emailAdd{tommaso.moretti@unina.it}
\emailAdd{noemi.frusciante@unina.it}

\title{Can cosmic voids ease the Hubble tension? Local expansion in $w_0w_a$CDM}

\abstract{
We investigate the effect of local cosmic voids on low-redshift measurements
of the Hubble rate in flat $w_0w_a$CDM cosmologies. Using a hydrodynamical
model for isolated spherical inverse top-hat underdensities, we compute the
void-induced Hubble shift as a function of redshift, enclosed density contrast $\delta_{\rm E}$
and cosmological parameters.
We find that the effect is mainly controlled by the matter sector, through
$\delta_{\rm E}$ and $\Omega_{\rm m,0}$, while dynamical dark energy gives only
subdominant late-time corrections. For a Planck-calibrated $\Lambda$CDM
background, matching the SH0ES value requires a present-day void with
$\delta_{\rm E}(z=0)\simeq -0.44$, substantially deeper than the KBC-like local
underdensity. A KBC-like void lowers the SH0ES--Planck discrepancy to about
$2\sigma$, but is not deep enough to fully reconcile the two measurements.
Allowing for dynamical dark energy, including regions motivated by recent DES
and DESI analyses, changes this result only at the percent level.
We also show that the reported redshift dependence of locally inferred $H_0$
values can be represented phenomenologically by an effective enclosed matter
profile. This reconstruction should be interpreted as a consistency test of
local-structure effects rather than as explanation for an evolving background value of $H_0$. Overall, local underdensities can affect low-redshift
Hubble-rate inferences, but they do not resolve the Hubble tension within the
spherical void setup considered here.
}

\begin{document}
\maketitle

\section{Introduction}

The determination of the Hubble constant, $H_0$, has become a consistency test of the cosmological model. Early universe observables constrain a tightly correlated parameter set which, in $\Lambda$CDM, implies a precise prediction for the present value of the expansion rate. Late universe measurements, in contrast, estimate $H_0$ more directly from the nearby distance–redshift relation. The two approaches are not in agreement: the SH0ES Cepheid–SN~Ia distance ladder measures $H_0 \simeq 73.17~{\rm km\,s^{-1}\,Mpc^{-1}}$~\cite{Riess:2021jrx,Breuval:2024lsv}, while \emph{Planck} Cosmic Microwave Background (CMB) data in flat $\Lambda$CDM prefer $H_0 \simeq 67.4~{\rm km\,s^{-1}\,Mpc^{-1}}$~\cite{Planck:2018vyg}. This mismatch is at the level of $\sim8\%$ and is often summarized as a $\gtrsim 5\sigma$ tension~\cite{CosmoVerseNetwork:2025alb}. Distance ladders based on the tip of the red giant branch typically yield intermediate values, reducing but not generically removing the discrepancy~\cite{Freedman:2019jwv}.

What makes this discrepancy consequential is that it stress-tests the pipeline that translates observations into a global expansion history. The early universe inference is model-dependent: CMB data tightly constrain the angular sound-horizon scale and physical densities, while $H_0$ emerges only after adopting a cosmological model, typically homogeneous and isotropic Friedmann--Lema\^{i}tre--Robertson--Walker (FLRW) cosmology with a specified dark energy (DE) sector. Late universe determinations are more direct but intrinsically local, and must contend with peculiar velocities, environmental astrophysics, and selection. The tension therefore invites a concrete question: is the expansion history inferred from the CMB being applied to our local patch in a systematically biased way?

Proposed solutions fall into three broad themes: unrecognized systematics; new physics that alters the pre-recombination calibration (often through the sound horizon); and late-time departures from minimal $\Lambda$CDM, including dynamical DE and violations of exact large-scale homogeneity. Late-time solutions face a tight net of constraints because baryon acoustic oscillations (BAO) and SNe~Ia jointly map the distance-redshift relation while large-scale structure (LSS) and lensing limit additional freedom~\cite{DiValentino:2021izs}. At the same time, the quality of low-redshift data makes the late-time sector a natural arena in which to test both DE dynamics and local-environment effects.

On the DE side, a widely used extension is the $w_0w_a$CDM, or Chevallier-Polarski-Linder (CPL), parameterization~\cite{Chevallier:2000qy,Linder:2002et},
which allows the DE equation of state (EoS) to evolve with time. Joint analyses combining BAO, supernova distances, and CMB information have renewed interest in this framework, partly because some dataset combinations prefer an evolving EoS different from that of a cosmological constant, $\Lambda$~\cite{DESI:2024mwx}. Recent multi-probe analyses further strengthen this motivation: using the full six-year Dark Energy Survey (DES) dataset, including supernovae, BAO, weak lensing and galaxy clustering, the DES Collaboration finds constraints that mildly prefer the region $w_0>-1$, $w_a<0$, with deviations from $\Lambda$CDM at the $2.3$--$3.2\sigma$ level depending on the probe combinations~\cite{DES:2026jmi}. The evidence is not uniform across datasets and depends sensitively on calibration, selection effects and probe combinations~\cite{Efstathiou:2024xcq}. This matters for the Hubble tension: any late-time modification that shifts inferred distances must remain consistent with the SN Hubble diagram, BAO measurements and structure-growth constraints across redshift~\cite{Lee:2022cyh,Tiwari:2026pzk}. This makes it important to ask whether late-time extensions of the background cosmology and local departures from homogeneity can be consistently disentangled. 

On the ``local environment'' side, if we reside in an underdense region (a local void), galaxies in and around it participate in a coherent outflow sourced by the void potential. If interpreted as purely cosmological redshift, this velocity field can mimic a higher local expansion rate. Before the tension sharpened, this possibility was often framed as a ``Hubble bubble'', motivating conservative low-$z$ cuts and flow corrections in SN analyzes~\cite{Zehavi:1998gz,Jha:2006fm}. The question now is quantitative: can local structure be large and deep enough to matter at the level implied by the tension?

A driver of this discussion is evidence for a sizable local underdensity on scales of a few hundred Mpc, often termed the ``Keenan-Barger-Cowie (KBC) void'' or ``Local Hole''. Near-infrared luminosity density measurements show a rise beyond $z\sim 0.07$, reaching values $\sim 1.5$ times higher than those inferred locally, suggestive of an underdensity within $\sim 300$~Mpc~\cite{Keenan:2013mfa}. Wide-area $K$-band counts and redshift distributions have been interpreted as a coherent $\sim 15-20\%$ underdensity on comparable scales~\cite{Whitbourn:2013mwa}. Cluster-based tracers provide a complementary probe: analyzes of X-ray-selected cluster samples report indications consistent with a local underdensity, with results depending on selection and tracer-to-mass mappings~\cite{Bohringer:2019tyj}. Together, these studies motivate the possibility that the nearby universe is not perfectly representative of the cosmic mean on the scales relevant for low-$z$ cosmography.

Connecting a ``Local Hole'' to $H_0$ requires dynamical modeling. How deep and extended must a void be to generate an apparent $\Delta H_0/H_0 \sim 0.08$, and is such a configuration compatible with structure formation and other observables? Much of the theoretical work addresses this using spherically symmetric inhomogeneous models, often of the Lema\^{i}tre--Tolman--Bondi (LTB) type~\cite{Garcia-Bellido:2008vdn,February:2009pv, Valkenburg:2012td,Redlich:2014gga}, or perturbative links between density contrast and peculiar velocities~\cite{Bernardeau:1992xxx,VanAcoleyen:2008cy}. In a KBC motivated setting, ref.~\cite{Hoscheit:2018nfl} showed that an LTB-like void can reduce the tension while remaining compatible with constraints on large-scale flows (including kinematic Sunyaev-Zel'dovich effect), but it does not typically eliminate it. A recurring conclusion is that resolving the full discrepancy requires voids that are uncomfortably large and/or extremely empty compared to what tracer measurements and $\Lambda$CDM statistics suggest.

Simulation-calibrated studies of cosmic variance in local $H_0$ determinations have reinforced this conclusion. Mock observers in large $N$-body volumes, analyzed with selections resembling nearby SN samples, typically yield percent-level scatter in locally inferred $H_0$. Bridging the full gap would require an unusually deep and extended underdensity, with effective contrasts approaching $\delta \sim -0.8$ on $\sim 100\,h^{-1}{\rm Mpc}$ scales in some characterizations~\cite{Huterer:2023ldv}. Related analyses using peculiar-velocity information and SN Hubble diagrams reach a similar conclusion: local structure is important to model, but is unlikely to be the sole explanation of the Hubble tension~\cite{Wu:2017fpr}.

Yet, the void hypothesis persists because the observational motivation is not obviously negligible and because mapping tracer densities to dynamical effects is subtle. Redshift-space distortions and coherent outflows can make the density contrast inferred from redshift surveys differ from the real-space matter contrast, and simple linear estimates may understate the impact of extended underdensities on low-$z$ inference~\cite{Haslbauer:2020xaa}. This invites stringent tests against BAO, which tightly constrains departures from the homogeneous distance-redshift relation. Ref.~\cite{Banik:2025dlo} confronts KBC-like void profiles with BAO measurements, emphasizing the role of the real-to-redshift-space mapping and the implied observer position within the void. Complementary analyses frame the problem as a global fit: void profiles that significantly affect $H_0$ must satisfy BAO distances, number counts and peculiar-velocity constraints, strongly restricting viable parameter space~\cite{Stiskalek:2025ibp}. Model-comparison studies likewise find that a void alone is insufficient as a complete resolution~\cite{Cai:2020tpy}, though more flexible reconstructions continue to be explored~\cite{Wang:2023reg}. In parallel, distance-ladder analyses emphasize null tests in which local-structure models are propagated through the inference pipeline; for example, ref.~\cite{Kenworthy:2019qwq} argue that local structure at the level suggested by current maps does not significantly bias SH0ES once flow corrections and redshift ranges are chosen to suppress such effects.

The intersection of a non-trivial local density environment and a possibly evolving DE sector therefore provides a well-defined setting in which to test whether local and background effects can be consistently separated in low-redshift measurements of the expansion rate. If evolving DE changes the inferred low-redshift distance scale, it can also modify the void-induced contribution needed to reconcile early- and late-time determinations of $H_0$. Conversely, if a local void perturbs the nearby Hubble diagram and the redshift distribution of low-$z$ SNe, it can feed back into constraints on $(w_0,w_a)$ obtained under the assumption of exact homogeneity~\cite{Banik:2026imu,Mazurenko:2024gwj}. Motivated by these considerations, we ask whether local expansion in cosmic voids within $w_0w_a$CDM can alleviate the discrepancy between early- and late-universe determinations of $H_0$. Thus, we study how the local density environment and the global expansion history jointly affect low-redshift inferences of the expansion rate using the hydrodynamical model for the non-linear evolution of spherical matter perturbations.

The hydrodynamical formalism has been extensively applied to model the spherical collapse of overdensities~\cite{Pace:2010sn,Abramo:2007iu,Bellini:2012qn,Frusciante:2020zfs} and was first extended to model the evolution of cosmic voids in $\Lambda$CDM, dark energy and modified gravity in our previous works~\cite{Moretti:2025gbp,Moretti:2026dfz,Moretti:2026axy}. This approach offers several advantages over the standard Newtonian one. In the latter, the non-linear evolution of spherically symmetric overdensities~\cite{Gunn:1972sv,Peebles:1980yev} or underdensities~\cite{Sheth:2003py,Demchenko:2016uzr,Massara:2018dqb} is described by tracking the evolution of the perturbation radius under Newtonian gravity, assuming a pressureless, self-gravitating fluid embedded in an expanding Universe. Firstly, it offers a simple and physically transparent description of the evolution of underdense regions by tracking the density contrast $\delta_{\rm E}$, while allowing the local Hubble parameter inside the void to be directly inferred from its evolution alone. Then, the framework can be naturally extended to modified gravity scenarios~\cite{Moretti:2026dfz}, with deviations from GR encoded through effective modifications of the gravitational potential $\mu_{\rm nl}$ entering the Poisson equation (see, e.g.,~\cite{Moretti:2026dfz,Moretti:2026axy}). Finally, it allows us to derive a cosmology-dependent mapping between the linear Lagrangian density contrast and its non-linear Eulerian counterpart, which is a fundamental ingredient in theoretical predictions of the Void Size Function~\cite{Sheth:2003py,Verza:2024rbm,Contarini:2026yfv}.

The paper is organized as follows. In section~\ref{sec:the_model}, we review the hydrodynamical model for isolated spherical inverse top-hat voids. In section~\ref{Sec:void_evolution_in_LCDM}, we discuss the evolution of the main void properties in the baseline $\Lambda$CDM cosmology. In section~\ref{sec:de_effects}, we study the void Hubble shift in flat $w_0w_a$CDM cosmologies. In section~\ref{Sec:implications_for_the_Hubble_tension}, we apply the framework to the Hubble tension, first in $\Lambda$CDM and then in dynamical DE backgrounds motivated by recent DES+DESI constraints. In section~\ref{Sec:redsfhit_dependence_of_the_locally_inferred_Hubble_rate}, we use the hydrodynamical model to investigate whether the reported redshift dependence of the locally inferred $H_0$ can be explained as an effective modulation induced by the enclosed matter distribution. We draw our conclusions in section~\ref{sec:conclusions}.

\section{The hydrodynamical model}\label{sec:the_model}

To study how cosmic voids affect the local expansion rate, we resort to the hydrodynamical framework introduced in~\cite{Moretti:2025gbp}. 
We adopt the Newtonian gauge for a perturbed spatially flat FLRW metric in cartesian coordinates,
\begin{align}
    {\rm d}s^2 = -\left(1+2\Psi\right){\rm d}t^2 + a^2(t)\left(1-2\Phi\right)\delta_{ij}\,{\rm d}x^i {\rm d}x^j \,,
    \label{eq:line_element_Newtonian_gauge}
\end{align}
where $a(t)$ is the scale factor of the universe with $t$ cosmic time, ${\bf x}$ is the comoving spatial coordinate vector, $\delta_{ij}$ is the three-dimensional Kronecker symbol, while $\Psi$ and $\Phi$ are the two gravitational potentials.

For the matter content, we neglect radiation and treat baryons and dark matter as a single pressureless fluid, while DE affects only the background expansion. We parameterize its EoS through the CPL form~\cite{Chevallier:2000qy,Linder:2002et},
\begin{align}
    w_{\rm DE}(a) \,=\, w_0+w_a(1-a)\,,
    \label{eq:EoS_DE}
\end{align}
where $w_0$ and $w_a$ are real constants, with $w_0$ denoting the present value of $w_{\rm DE}$ and $w_a$ corresponding to minus its derivative with respect to $a$ evaluated today.

To obtain the dynamics of matter perturbations, we combine the Einstein field equations (EFE) together with the conservation of the stress-energy tensor, in the Newtonian, weak-field, sub-horizon regime. From the EFE, the gravitational potentials satisfy
\begin{align}
    \nabla_{\mathbf{x}}^2\Psi &= 4\pi G a^2 \bar{\rho}_{\rm m}\,\delta_{\rm E}\,,
    \label{eq:Poisson_equation}\\
    \nabla_{\mathbf{x}}^2\left(\Phi+\Psi\right) &= 8\pi G a^2 \bar{\rho}_{\rm m}\,\delta_{\rm E}\,,
    \label{eq:lensing_equation}
\end{align}
where $\bar{\rho}_{\rm m}$ is the background matter density, $G$ is the Newtonian gravitational constant, and $\delta_{\rm E}$ is the Eulerian (fully non-linear) matter density contrast. From eqs.~\eqref{eq:Poisson_equation} and~\eqref{eq:lensing_equation}, we have $\Phi=\Psi$, which comes from the absence of anisotropic stresses. The conservation equations, i.e. $\nabla_{\mu}T^{\mu\nu}=0$, then provide the continuity and Euler equations, i.e.
\begin{align}
    \frac{\partial\delta_{\rm E}}{\partial t} + (1+\delta_{\rm E})\,\nabla_{\mathbf{x}}\cdot\vec{u} &= 0\,,
    \label{eq:euler_equation}
    \\
    \frac{\partial\vec{u}}{\partial t} + 2H\vec{u} + (\vec{u}\cdot\nabla_{\mathbf{x}})\vec{u} + \frac{1}{a^2}\nabla_{\mathbf{x}}\Psi &= 0\,,
    \label{eq:continuity_equation}
\end{align}
where $\vec{u}$ is the spatial component of the comoving peculiar velocity and $H\equiv ({\rm d} a/{\rm d}t)/a$. Assuming spherical symmetry, these equations can be combined into a single fully non-linear evolution equation for the density contrast,
\begin{align}
    \delta_{\rm E}'' + \left(2+\frac{H'}{H}\right)\delta_{\rm E}'
    -\frac{4}{3}\frac{(\delta_{\rm E}')^2}{1+\delta_{\rm E}}
    -\frac{3}{2}\Omega_{\rm m}\,(1+\delta_{\rm E})\,\delta_{\rm E}
    =0\,,
    \label{eq:non_linear_evolution_equation}
\end{align}
where primes denote derivatives with respect to $\ln a$ and $\Omega_{\rm m}=\bar\rho_{\rm m}/3M_{\rm pl}^2H^2$ is the background matter fraction with $M_{\rm pl}^2=(8\pi G)^{-1}$. This equation is the basic dynamical ingredient of the hydrodynamical model for cosmic void evolution. In the form of eq.~\eqref{eq:non_linear_evolution_equation}, the evolution depends on the background expansion history through $\Omega_{\rm m}$ and $H'/H$, but it is independent of the absolute normalization $H_0$. 

We now apply eq.~\eqref{eq:non_linear_evolution_equation} to describe the dynamics of an isolated spherical inverse top-hat void. This requires specifying the density profile and the initial conditions (ICs). We take
\begin{align}
    \delta_{\rm E}(t_{\rm in},r_{\rm in}) \,=\,
    \begin{cases}
    \delta_{\rm v,in} & \text{for } r_{\rm in} \leq r_{\rm v,in} \\
    0 & \text{for } r_{\rm in} > r_{\rm v,in}
    \end{cases}\,,
    \label{Eq:initial_top_hat_delta}
\end{align}
where $t_{\rm in}$ is the initial time, $r_{\rm in}=a(t_{\rm in})\,x$ is the initial physical radius, $r_{\rm v,in}$ is the initial void radius, and $\delta_{\rm v,in}<0$ is the initial density contrast inside the void. Within this setup, the density profile preserves its inverse top-hat form throughout the evolution, up to shell-crossing, where the single-stream hydrodynamical description ceases to apply~\cite{Moretti:2025gbp,Moretti:2026dfz}. This follows from the fact that all fluid elements inside the void obey eq.~\eqref{eq:non_linear_evolution_equation} with the same ICs. The dynamics can therefore be reconstructed by following a single representative point.
In practice, we integrate eq.~\eqref{eq:non_linear_evolution_equation} for this representative point. We start the integration at $a_{\rm in}=10^{-7}$ and, since radiation is neglected, this corresponds to an idealized early-time EdS regime. This choice is not meant to provide a realistic description of the radiation era, but only to fix the ICs in a matter-dominated limit. The late-time results are insensitive to this assumption, as discussed in appendix B of~\cite{Moretti:2025gbp}.

Since eq.~\eqref{eq:non_linear_evolution_equation} is a second-order ordinary differential equation, two ICs must be specified. One condition is obtained by requiring that the void is initially in the linear regime, $\delta_{\rm E}=\delta_{\rm L}$, and by neglecting the decaying mode. In the matter-dominated limit, these assumptions give
\begin{align}
    \delta_{\rm E}'(a_{\rm in})=\delta_{\rm E}(a_{\rm in})\equiv \delta_{\rm v,in}\,,
    \label{Eq:initial_condition_delta_prime}
\end{align}
which fixes the initial derivative in terms of the initial density contrast, but does not determine the value of $\delta_{\rm v,in}$ itself. This value is instead fixed by imposing a late-time condition. For instance, to model a void with $\delta_{\rm E}(z=0)=-0.5$, we tune $\delta_{\rm v,in}$ through a shooting procedure until the evolved solution reaches this value. Once $\delta_{\rm v,in}$ is fixed, the full void dynamics is determined. This construction remains valid only before shell-crossing, namely before the outermost shell of the void, which we define as the \textit{void radius}, reaches the surrounding environment. The shell-crossing threshold $\delta_{\rm E,sc}(z)$ is determined by the following condition~\cite{Moretti:2025gbp}
\begin{align}
    \frac{\mathrm{d} \delta_{\rm E}}{\mathrm{d} \delta_{\rm v,in}}
    =
    -\frac{1+\delta_{\rm E}}
    {(1+\delta_{\rm v,in})\,\delta_{\rm v,in}}\,.
    \label{Eq:Shell_crossing_Delta}
\end{align}
Beyond this point, the single-stream hydrodynamical description adopted here ceases to apply. 

Additionally, we clarify the relation between the top-hat description used here and the more general spherical-shell formulation. In spherical symmetry, the dynamics of a shell of radius $R$ is controlled by the enclosed mass, or equivalently by the averaged density contrast
\begin{align}
    \Delta_{\rm E}(R,t)
    =
    \frac{3}{4\pi R^{3}}
    \int_0^R{\rm d}s\,s^{2}
    \int{\rm d}\Omega\,\delta_{\rm E}(\mathbf{s},t)\,,
    \label{eq:mean_density_contrast}
\end{align}
where $s=\lvert\mathbf{s}\rvert$ and $\Omega$ is the solid angle. A shell enclosing a given value of $\Delta_{\rm E}$ obeys the same equation of motion as the outer shell of an inverse top-hat void with the same averaged density contrast (see~\cite{Moretti:2025gbp}). For the top-hat configuration considered here, $\delta_{\rm E}$ is spatially constant inside the void, so $\Delta_{\rm E}=\delta_{\rm E}$. The local and averaged descriptions are therefore equivalent for the dynamics studied in this work.

The hydrodynamical solution for $\delta_{\rm E}$ also allows us to reconstruct the evolution of the void radius. To do so, we impose mass conservation for spherical shells. Denoting by $R_{\rm v}(t)$ the physical radius of the outermost shell of the inverse top-hat, the enclosed mass is conserved according to
\begin{align}
    M
    \,=\,
    \frac{4\pi}{3}\,\bar{\rho}_{\rm m}(t)\,
    \left[1+\delta_{\rm E}(t)\right]\,R_{\rm v}^3(t)
    \,=\,
    \frac{4\pi}{3}\,\bar{\rho}_{\rm m}(t_{\rm in})\,
    \left[1+\delta_{\rm v,in}\right]\,r_{\rm v,in}^3\,.
    \label{eq:mass_conservation}
\end{align}
Thus, once $\delta_{\rm E}(t)$ is known, eq.~\eqref{eq:mass_conservation} determines the evolution of the ratio $R_{\rm v}(t)/r_{\rm v,in}$. Thus, the absolute normalization of the radius is not fixed. In analogy with the discussion for $\delta_{\rm E}$, it can be set by imposing a late-time condition, e.g. $R_{\rm v}(z=0)=100\,{\rm Mpc}/h$, which fixes $r_{\rm v,in}$. In this case, however, no shooting procedure is required: since the dynamics is invariant under a constant rescaling of the radius, $r_{\rm v,in}$ is obtained by a simple rescaling of the solution.

Then, we can also compute the local expansion rate inside the void as
\begin{align}
    H_{\rm v}
    \equiv
    \frac{\dot{R}_{\rm v}}{R_{\rm v}}
    =
    H\,\frac{R_{\rm v}'}{R_{\rm v}}
    \equiv
    \mathcal{H}_{\rm v} H\,,
    \label{eq:Hv_def}
\end{align}
where $\mathcal{H}_{\rm v}$ is the local expansion rate normalized to the background one. We stress that obtaining the absolute value of $H_{\rm v}$ requires fixing the normalization of the background expansion, $H_0$. If, instead, one is interested only in the fractional deviation from the background, this normalization cancels out. Using mass conservation, $\mathcal{H}_{\rm v}$ can be written directly in terms of the density contrast as
\begin{align}
    \mathcal{H}_{\rm v}
    =
    1
    -
    \frac{1}{3}
    \left(
    \frac{\delta_{\rm E}'}{1+\delta_{\rm E}}
    \right)\,.
    \label{eq:Hv_delta_relation}
\end{align}
from which it is explicit that the dependence of $\mathcal{H}_{\rm v}$, and consequently of $H_{\rm v}$, on the void properties is entirely encoded in the density evolution, with no dependence on the absolute void size.

\section{Results}\label{sec:results}
In this section, we assess how dynamical DE of the CPL form affects the local expansion rate inside cosmic voids, or equivalently the Hubble parameter that would be inferred by an observer performing measurements within an underdense region. 
The discussion is organized around the following points:
\begin{itemize}
    \item We first review the evolution of an isolated inverse top-hat void in the baseline $\Lambda$CDM cosmology. This establishes the reference behaviour of the density contrast, radius excess, local expansion-rate excess, and acceleration with respect to the homogeneous background.
    \item We then show that DE leaves a characteristic imprint on the redshift evolution of a single void Hubble shift, $\Delta H_{\rm v}$, with the latter we mean the relative difference between the Hubble rate in a void and the background one (see eq.~\eqref{Eq:def_three_quantities_percentage_difference}). For a fixed initial void trajectory, DE suppresses the matter-driven evacuation of the void at low redshift, producing a turnover in $\Delta H_{\rm v}$ which is absent in the EdS model.
    \item We quantify the dependence of $\Delta H_{\rm v}$ on the redshift and the void depth and on the cosmological parameters. We place particular emphasis on its present-day value, i.e. $\Delta H_{\rm v}(z=0)$, showing that its dominant dependence is on the density contrast $\delta_{\rm E}$ and on the present-day background matter fraction, $\Omega_{\rm m,0}$, while $w_0$ produces a smaller correction and $w_a$ gives a subdominant modulation over the parameter range considered.
    \item We isolate the residual CPL correction relative to $\Lambda$CDM and show that it acts mainly as an overall rescaling of $\Delta H_{\rm v}(z=0)$. The effect of $w_a$ is not universal, but depends on the chosen value of $w_0$.
    \item We apply the framework to the Hubble tension problem in both $\Lambda$CDM and dynamical DE backgrounds. For each cosmology, we determine the present-day void depth required to reproduce the SH0ES value through the void-induced local expansion rate, fixing the background normalization to $H_0^{\rm P18}=67.4\,{\rm km\,s^{-1}\,Mpc^{-1}}$. We use this matching to quantify how the required $\delta_{\rm E}$ depends on cosmological parameters.
    \item In $\Lambda$CDM, we quantify the residual statistical tension as a function of the assumed present-day void depth, fixing both $\Omega_{\rm m,0}$ and $H_0$ to their Planck values. This allows us to assess whether KBC-like underdensities can remove the discrepancy between SH0ES and Planck or only alleviate it.
    \item We then apply the same machinery to recent DES+DESI constraints on dynamical DE~\cite{DESI:2025zgx,DES:2026jmi}, in order to assess the combined implications of these constraints and local void effects for the Hubble tension.
    \item Finally, we use the hydrodynamical model to show that the reported redshift dependence of the locally inferred $H_0$~\cite{Jia:2022ycc,Jia:2024wix} can be phenomenologically represented, at the bin-by-bin level, by an effective enclosed matter profile, and we discuss whether the required profile is physically plausible and compatible with observations of the local matter distribution.
\end{itemize}
In sections~\ref{Sec:void_evolution_in_LCDM} and~\ref{sec:de_effects}, we take as baseline model the flat $\Lambda$CDM cosmology
\begin{align}
    \Lambda{\rm CDM}
    =
    \left\{
    \Omega_{\rm m,0}=0.32\,,
    w_0=-1\,,
    w_a=0
    \right\}\,.
    \label{eq:baseline_cosmology}
\end{align}
When varying $\Omega_{\rm m,0}$, $w_0$, or $w_a$ around the baseline model in order to isolate the dependence of a given quantity on the cosmological parameters, we use broad ranges,
\begin{align}
    \Omega_{\rm m,0}\in[0.05,1.0]\,,\quad
    w_0\in[-2.0,-0.4]\,,\quad
    w_a\in[-2.0,0.1]\,.
\end{align}
These intervals include values that lie outside current observationally allowed regions. They should therefore be interpreted only as diagnostic scans.
However, when comparing with observationally motivated cosmologies as in section~\ref{Sec:implications_for_the_Hubble_tension}, instead, we use parameter values tied to specific datasets. For $\Lambda$CDM, we adopt the Planck-calibrated values
\begin{align}
    H_0^{\rm P18}=67.4\,{\rm km\,s^{-1}\,Mpc^{-1}}\,,
    \quad
    \Omega_{\rm m,0}^{\rm P18}=0.315\,.
\end{align}
For dynamical DE, we consider the $1\sigma$ region motivated by recent DES+DESI constraints, which mildly prefer $w_0>-1$ and $w_a<0$. In particular, for the late-time DES+DESI combination we use
\begin{align}
    \Omega_{\rm m,0}=0.31\,,\quad
    w_0=-0.84^{+0.06}_{-0.07}\,,\quad
    w_a=-0.53^{+0.33}_{-0.28}\,,
\end{align}
while for the DES+DESI+CMB combination we use
\begin{align}
    \Omega_{\rm m,0}=0.312\,,\quad
    H_0=67.4\,{\rm km\,s^{-1}\,Mpc^{-1}}\,,
    \quad
    w_0=-0.82\pm0.05\,,\quad
    w_a=-0.63^{+0.21}_{-0.18}\,.
\end{align}


\subsection{Void evolution in \texorpdfstring{$\Lambda$CDM}{LambdaCDM}}
\label{Sec:void_evolution_in_LCDM}

Before studying the impact of dynamical DE, we first review the basic evolution of an isolated spherical inverse top-hat void in the baseline flat $\Lambda$CDM model. This provides the physical reference against which the $w_0w_a$CDM results will be interpreted.

\begin{figure}[t]
    \centering
    \includegraphics[width=1.0\linewidth]{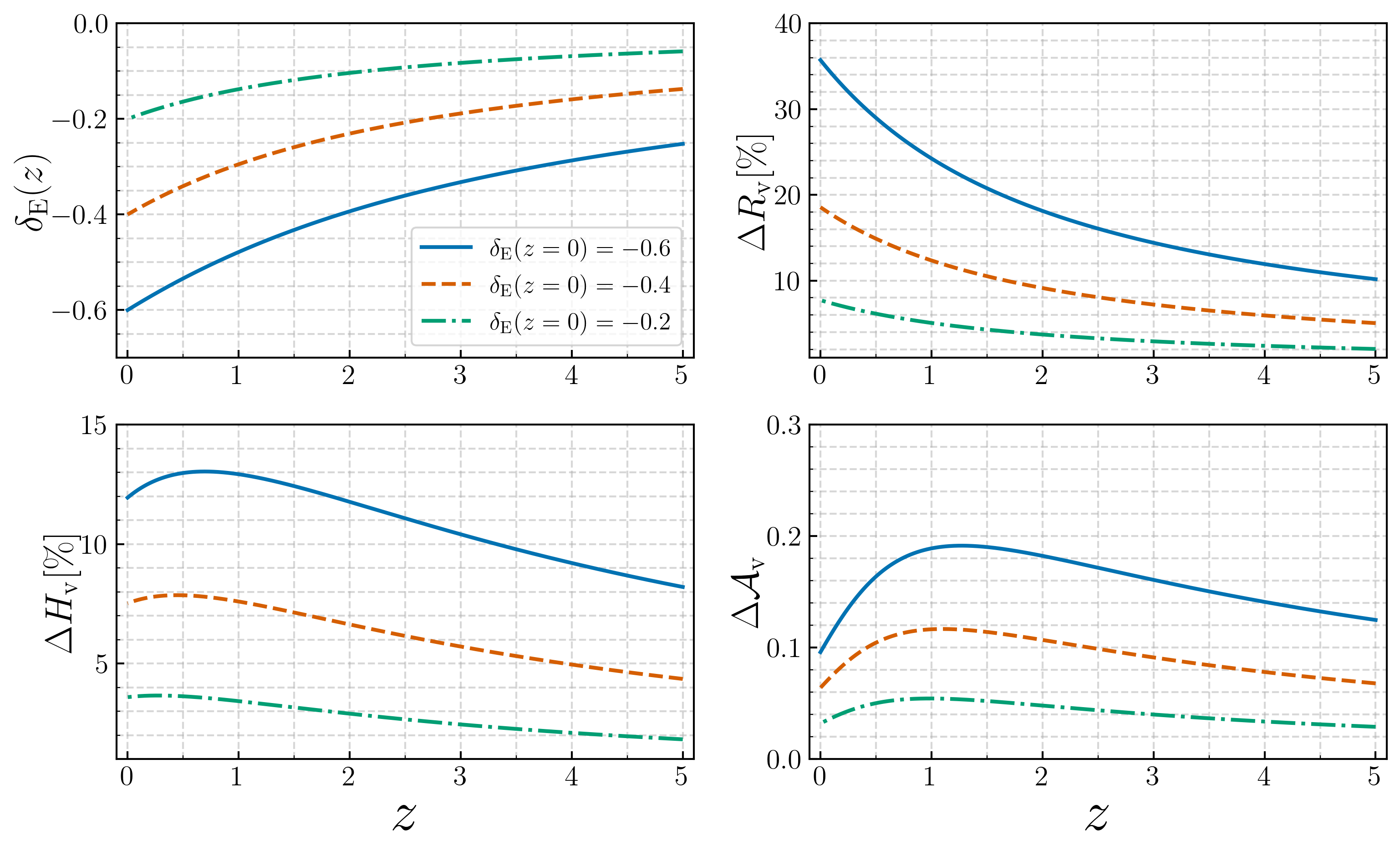}
    \caption{
    Evolution of an isolated spherical inverse top-hat void in the baseline $\Lambda$CDM cosmology. 
    The curves correspond to present-day void depths $\delta_{\rm E}(z=0)=\{-0.6,-0.4,-0.2\}$. 
    The panels show the evolution with redshift of the density contrast (top left), the percentage relative radius excess (top right), the percentage local expansion-rate excess (bottom left), 
    and the acceleration difference with respect to the homogeneous background (bottom right). 
    }
    \label{fig:four_quantities_void_evolution}
\end{figure}

In figure~\ref{fig:four_quantities_void_evolution}, we compare the evolution of the outermost void shell with that of the corresponding background shell, namely the shell that would follow the homogeneous expansion if the initial underdensity were absent. The void shell is described by its physical radius $R_{\rm v}(t)$, while the background shell simply follows $R\propto a(t)$. We set $R_{\rm v}(t_{\rm in})=a_{\rm in}$, so that the two evolutions start from the same initial radius and can be compared directly. The figure spans the redshift interval $z\in[0,5]$ and includes three representative cases with $\delta_{\rm E}(z=0)=\{-0.6,-0.4,-0.2\}$, corresponding to the underdensities typically found in both $N$-body simulations and observations (see, e.g.,~\cite{Verza:2019tvg,Bayer:2021iyb,Verza:2024rbm}). We show $\delta_{\rm E}$ (upper left panel), which directly quantifies the matter deficit inside the void relative to the background, the relative radius difference (upper right panel), the local expansion rate excess (lower left panel), and the acceleration difference (lower right panel). The latter three quantities are defined respectively as
\begin{align}
    \Delta R_{\rm v}[\%] \equiv 100 \times \left(\frac{R_{\rm v}-a}{a}\right)\,, \quad
    \Delta H_{\rm v}[\%] \equiv 100 \times \left(\frac{H_{\rm v}-H}{H}\right)\,, \quad
    \Delta \mathcal{A}_{\rm v} \equiv \left(\frac{\ddot{R}{\rm v}}{R_{\rm v}} - \frac{\ddot{a}}{a}\right)\,.
    \label{Eq:def_three_quantities_percentage_difference}
\end{align}
Since $\ddot{a}/a$ crosses zero, the last quantity is shown as a direct difference rather than as a relative percentage.

The evolution follows the standard physical picture of void formation. As time increases, matter is evacuated from the underdense region, so that $\delta_{\rm E}$ becomes increasingly negative. The outer shell therefore expands faster than the corresponding background shell, producing a monotonically growing radius excess $\Delta R_{\rm v}[\% ]$. Deeper voids show a larger departure from the background at all redshifts.
The most relevant quantity for the following analysis is the local expansion-rate excess $\Delta H_{\rm v}$. Eqs.~\eqref{eq:mass_conservation} and~\eqref{eq:Hv_delta_relation} give
\begin{align}
    \Delta H_{\rm v}
    = \frac{{\rm d}\ln R_{\rm v}}{{\rm d}\ln a}-1
    = -\frac{1}{3}\frac{{\rm d}\ln(1+\delta_{\rm E})}{{\rm d}\ln a}\,.
\end{align}
Thus, $\Delta H_{\rm v}$ directly measures the void formation rate: it quantifies how rapidly the void expands relative to the background, or equivalently how rapidly the density contrast deepens. In $\Lambda$CDM, $\Delta H_{\rm v}$ increases while void formation is efficient during matter domination, reaches a broad maximum at intermediate redshift, and then decreases toward $z=0$ as $\Lambda$ suppresses further growth. This provides a clear illustration of how $\Lambda$ slows down the growth of cosmic structures. The turnover is therefore the key baseline signature against which the effect of dynamical DE will be compared. 

Finally, the acceleration difference is strictly positive over the whole redshift range, showing that the void shell always accelerates more than the corresponding background shell. Its amplitude also increases for deeper voids, confirming that more underdense regions depart more strongly from the background dynamics.

\subsection{Dark energy effects on the local expansion rate}
\label{sec:de_effects}

We now ask how sensitive the local expansion rate of a cosmic void is to the assumed DE sector. Using the spherical void dynamics described above, we quantify how changes in the background expansion history modify the local rate $H_{\rm v}$ defined in eq.~\eqref{eq:Hv_def}, and hence its departure from the homogeneous Hubble rate.

We first follow the redshift evolution of a single void trajectory. The ICs are imposed at early times, when all the cosmologies considered here approach the same matter-dominated limit. We fix the initial density contrast by requiring that the corresponding EdS evolution reaches $\delta_{\rm E}(z=0)=-0.4$, a representative depth for cosmic voids identified in simulations and observations. The same ICs are then evolved in different DE cosmologies. This construction isolates the effect of the background expansion history on void dynamics.
\begin{figure}[t]
    \centering
    \includegraphics[width=1.0\linewidth]{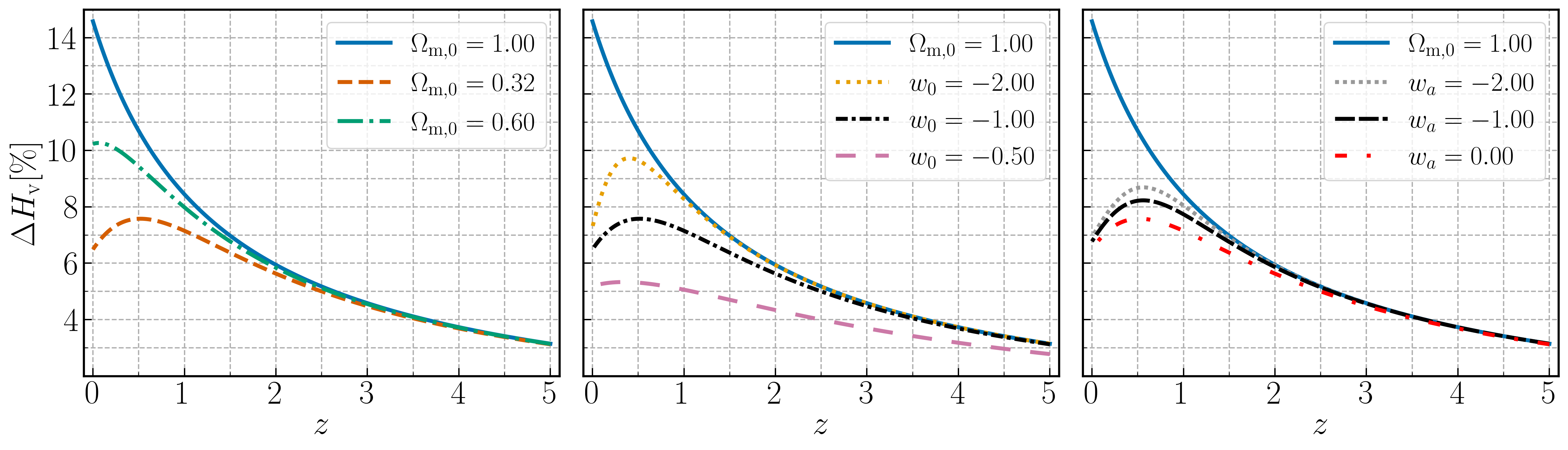}
    \caption{Evolution of $\Delta H_{\rm v}[\%]$ over the range $z\in[0,5]$ for a void whose ICs are fixed by tuning the initial density contrast so that the corresponding EdS solution reaches $\delta_{\rm E}(z=0)=-0.4$. The same ICs are then evolved in different CPL cosmologies. From left to right, we vary $\Omega_{\rm m,0}=\{0.32,0.6,1\}$, $w_0=\{-2,-1,-0.5\}$, and $w_a=\{-2,-1,0\}$, with all other parameters fixed to the baseline $\Lambda$CDM model of eq.~\eqref{eq:baseline_cosmology}. The EdS solution is shown as a reference in each panel.}
    \label{fig:hubble_parameter_voids_evolution}
\end{figure}

Figure~\ref{fig:hubble_parameter_voids_evolution} shows the redshift evolution of $\Delta H_{\rm v}[\%]$, defined in eq.~\eqref{Eq:def_three_quantities_percentage_difference}, for this single evolving void over the range $z\in[0,5]$. The three panels vary, respectively, $\Omega_{\rm m,0}=\{0.32,0.6,1\}$, $w_0=\{-2,-1,-0.5\}$, and $w_a=\{-2,-1,0\}$ around the baseline $\Lambda$CDM model of eq.~\eqref{eq:baseline_cosmology}. Since $\Delta H_{\rm v}$ depends only on the ratio $H_{\rm v}/H$, the absolute normalization of $H_0$ is irrelevant for this comparison.

Before discussing the dependence on the DE parameters, it is useful to recall the EdS behavior. In a purely matter-dominated universe, $\Delta H_{\rm v}[\%]$ grows monotonically toward low redshift, or equivalently as the void deepens. As $\delta_{\rm E}$ becomes more negative, the outer shell is increasingly less decelerated relative to the corresponding background shell, making the relative expansion of the void more efficient. This behavior is a generic property of EdS void evolution and is not tied to the specific final depth chosen. It can also be derived from the analytical EdS solution; see e.g.~\cite{Sheth:2003py}.

\begin{figure}[t]  
    \centering\includegraphics[width=0.8\textwidth]{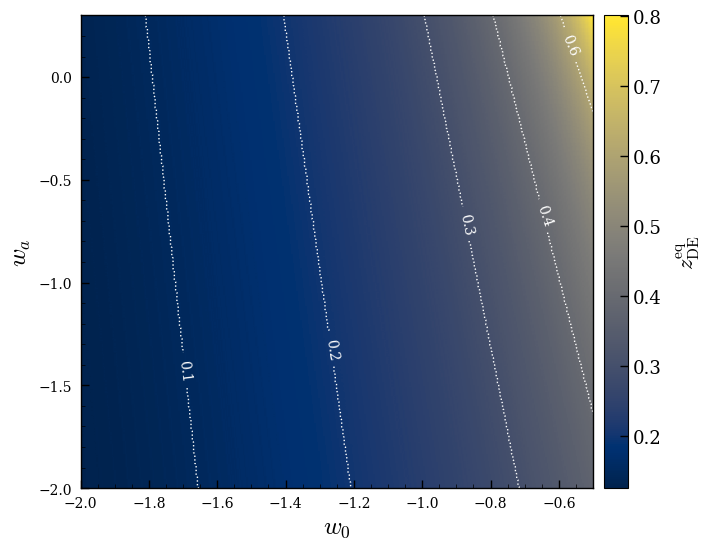}
    \caption{The redshift of matter–dark energy equality, $z^{\rm eq}_{\rm DE}$ as a function of $w_0$ and $w_a$, while keeping $\Omega_{\rm m,0} = 0.32$ fixed.}
    \label{Fig:z_eq}
\end{figure}
Once DE is included, the curves display two immediate features. First, the deviations from EdS appear only at low redshift, while at high redshift all models converge to the EdS limit, where DE is dynamically negligible. Second, the DE curves lie below the EdS prediction, showing that DE suppresses the expansion-rate excess inside the void. At the level of the hydrodynamical equation, this suppression is encoded in eq.~\eqref{eq:non_linear_evolution_equation}: as DE becomes dynamically relevant, $\Omega_{\rm m}$ decreases and the matter source driving void evacuation becomes less efficient.
Finally, the most distinctive imprint of DE is the appearance of a maximum in $\Delta H_{\rm v}$. Before this maximum, matter still drives the evacuation of the void and the local expansion excess continues to grow, although less efficiently than in EdS. After the maximum, DE suppression dominates and $\Delta H_{\rm v}$ decreases. In the asymptotic DE-dominated regime, the matter source becomes negligible and the local and background expansion rates converge, with $H_{\rm v}/H\to 1$ and $\Delta H_{\rm v}\to 0$.

We stress that the redshift of the maximum, $z_{\rm max}$, should not be identified with the matter--DE equality redshift. The latter is a property of the homogeneous background, whereas $z_{\rm max}$ also depends on the void trajectory selected by the ICs, and therefore on the rate at which the underdensity evolves.

Finally, the trends with cosmological parameters trace the onset of DE domination (see figure~\ref{Fig:z_eq}). Decreasing $\Omega_{\rm m,0}$, or increasing $w_0$ or $w_a$ within the CPL range considered here, makes DE dynamically relevant at earlier times. This weakens the matter-driven evacuation of the void, enhances the suppression of $\Delta H_{\rm v}$ relative to EdS, and shifts its maximum to higher redshift. Models in which matter domination lasts longer instead remain closer to the EdS limit and exhibit a delayed maximum. Within the parameter range shown, the dominant variations are associated with $\Omega_{\rm m,0}$ and $w_0$, whereas $w_a$ provides only a milder modulation.
 
What we discussed until now is useful to compute a more directly accessible quantity with observations: the local expansion-rate excess associated with voids of a given depth at a fixed redshift. We therefore consider the local expansion rate, or equivalently its fractional excess, as a function of the void depth and redshift, namely $H_{\rm v}=H_{\rm v}(\delta_{\rm E},z)$. In practice, for each pair $(\delta_{\rm E},z)$, we use the hydrodynamical framework to select the void trajectory that reaches the chosen depth at that redshift, and then compute the corresponding value of $H_{\rm v}$. This construction quantifies how the local expansion rate responds to the depth of voids at a given epoch. 
 
\begin{figure}[t]
    \centering
    \includegraphics[width=1.0\linewidth]{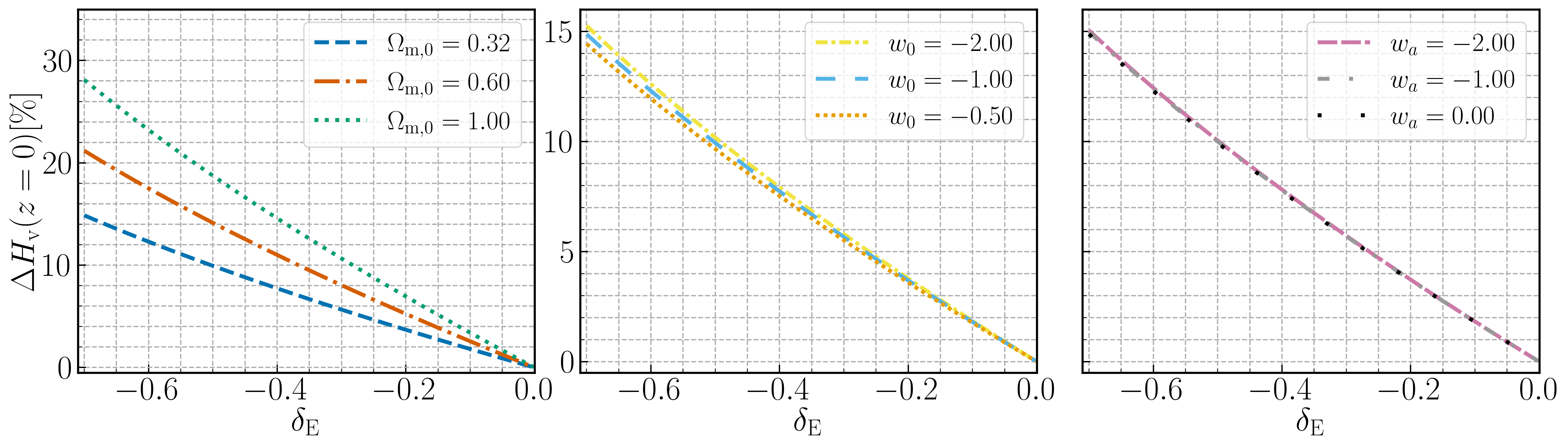}
    \caption{Present-day void Hubble shift, $\Delta H_{\rm v}(\delta_{\rm E},z=0)$, as a function of the
    density contrast $\delta_{\rm E}$. From left to right, the
    panels show variations of $\Omega_{\rm m,0}$, $w_0$, and $w_a$, with all other parameters kept
    fixed to their values in the baseline model.} 
    \label{fig:hubble_parameter_z=0}
\end{figure}

To study the impact of DE parameters on $H_{\rm v}(\delta_{\rm E},z)$, one must fix either redshift or void depth. We begin with the former case in figure~\ref{fig:hubble_parameter_z=0} where we display $\Delta H_{\rm v}(\delta_{\rm E},z=0)[\%]$ as a function of $\delta_{\rm E}$, over the range $\delta_{\rm E}\in[-0.7,-0.001]$. From left to right, the panels show the effect of varying $\Omega_{\rm m,0}$, $w_0$, and $w_a$, respectively, around the fiducial model. All displayed values of $\delta_{\rm E}$ lie above the shell-crossing threshold, $\delta_{\rm E,sc}$, below which the model ceases to provide a physically meaningful description~\cite{Moretti:2025gbp}. 

Figure~\ref{fig:hubble_parameter_z=0} shows that $\Delta H_{\rm v}(z=0)$ is mainly controlled by the void depth.
For all cosmologies considered, the local expansion-rate excess increases monotonically
as $\delta_{\rm E}$ becomes more negative and vanishes in the homogeneous
limit. At fixed $\delta_{\rm E}$, the cosmological dependence follows a
clear hierarchy: $\Omega_{\rm m,0}$ gives the dominant contribution, $w_0$
induces a smaller but visible modulation, and $w_a$ has only a subdominant
effect over the range explored. This indicates that, at $z=0$, the efficiency of
void evacuation is primarily set by the matter fraction, while the detailed CPL
evolution of the DE sector provides a secondary correction.
\begin{figure}[t]
    \centering
    \includegraphics[width=1.0\linewidth]{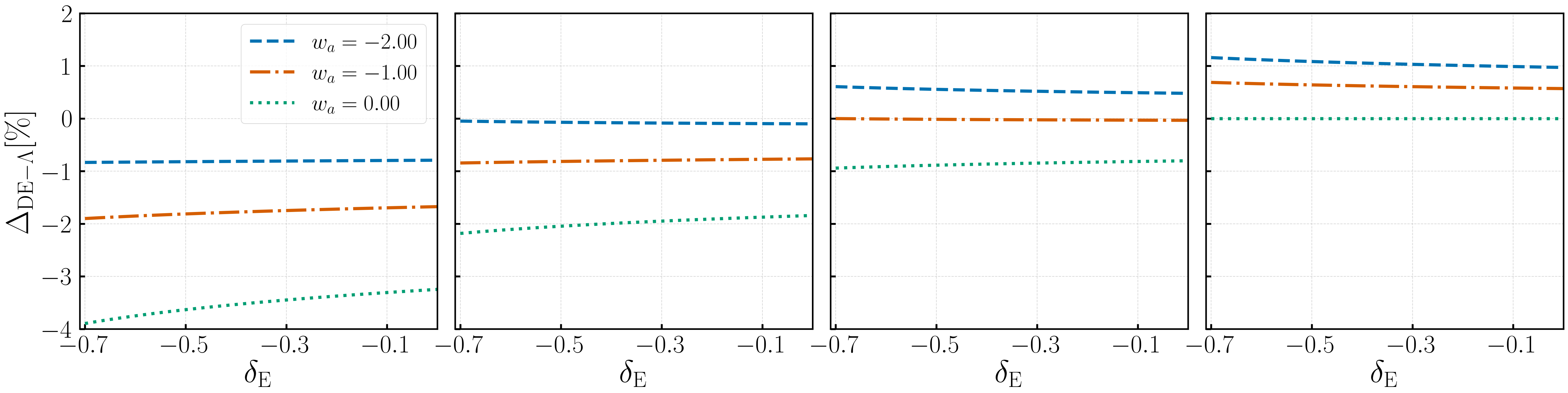}
    \caption{
    Relative change of the present-day void Hubble shift with respect to the
    baseline $\Lambda$CDM model as defined in eq.~\eqref{eq:baseline_cosmology} as a function of the density
    contrast $\delta_{\rm E}$. From left to right, the panels show
    fixed values of $w_0=\{-0.4,-0.6,-0.8,-1.0\}$, while the curves in each panel
    correspond to different values of $w_a$. The matter density is fixed to
    $\Omega_{\rm m,0}=0.32$.
     }
    \label{fig:perc_diff_hubble_z=0}
\end{figure}

The very weak dependence on $w_a$ found in figure~\ref{fig:hubble_parameter_z=0} does not, by itself, imply that its effect is universally negligible, since the response to $w_a$ may change once the underlying value of $w_0$ is varied. Therefore, in figure~\ref{fig:perc_diff_hubble_z=0}, we show the relative variation of
$\Delta H_{\rm v}(\delta_{\rm E},z=0)$ of the CPL model with respect to the baseline $\Lambda$CDM prediction:
\begin{align}
    \Delta_{\rm DE - \Lambda}[\%]
    \,=\,
    100\,\times \left[
    \frac{
    \left(\Delta H_{\rm v}\right)_{{\rm DE}}
    -
    \left(\Delta H_{\rm v}\right)_{\Lambda}
    }{
    \left(\Delta H_{\rm v}\right)_{\Lambda}
    }\right]\,,
\end{align}
as a function of $\delta_{\rm E}$ over the range $\delta_{\rm E}\in[-0.7,-0.001]$, for fixed $\Omega_{\rm m,0}=0.32$, with four panels corresponding from left to right to $w_0=\{-0.4\,,-0.6\,,\,-0.8\,,-1.0\}$, and with $w_a=\{-2\,,-1\,,0\}$ varied within each panel. 
This representation removes the dominant dependence on the absolute size of the
void Hubble shift and highlights the residual effect of the CPL parameters. The
curves are almost independent of $\delta_{\rm E}$, showing that the CPL
correction acts mainly as an overall rescaling rather than as a modification of
the depth dependence. At fixed $w_0$, changing $w_a$ shifts the curves vertically, with a larger effect
for less negative values of $w_0$, consistently with the way these parameters
modify the redshift of matter-DE equality. Thus, the impact of $w_a$ is not universal:
it depends on the value of $w_0$. 

\begin{figure}[t]
    \centering
    \includegraphics[width=1.0\linewidth]{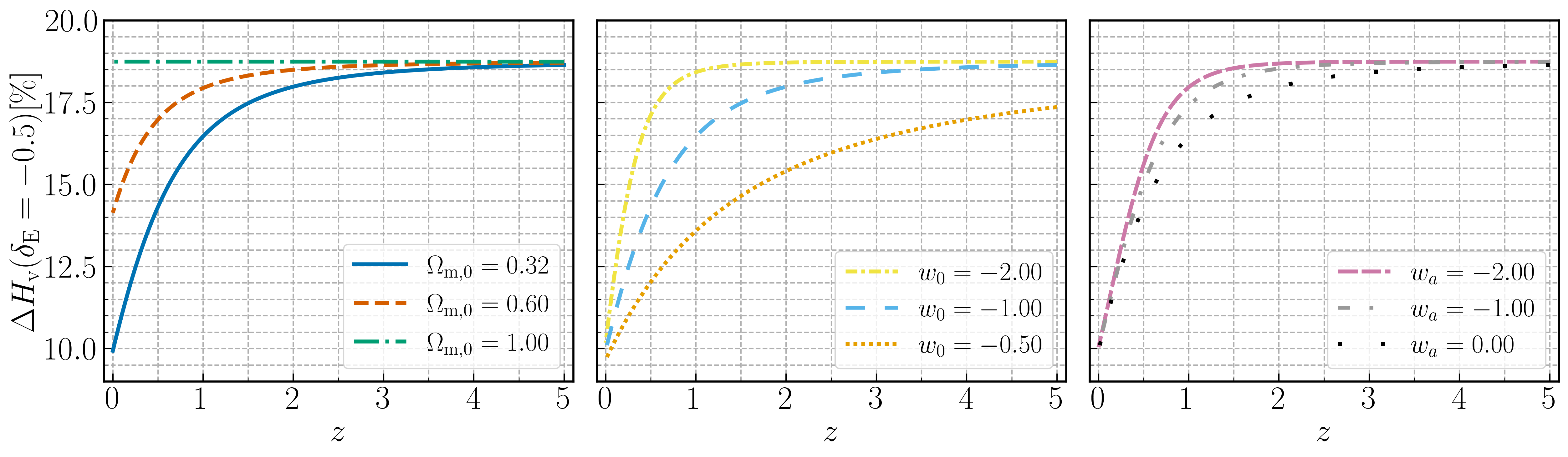}
    \caption{Redshift evolution of the void Hubble shift, 
    $\Delta H_{\rm v}(\delta_{\rm E}=-0.5,z)$, at fixed density contrast. 
    From left to right, the panels illustrate the effect of varying $\Omega_{\rm m,0}$, 
    $w_0$, and $w_a$, respectively, while keeping all other parameters fixed to their 
    values in the baseline model.}
    \label{fig:hubble_parameter_d_fixed}
\end{figure}

Finally, in figure~\ref{fig:hubble_parameter_d_fixed} we show $\Delta H_{\rm v}(\delta_{\rm E}=-0.5,z)[\%]$ as a function of the redshift, over the range $z\in\{0,5\}$. The figure is organized as figure~\ref{fig:perc_diff_hubble_z=0}. In EdS, the curve is independent of redshift, reflecting the self-similar voids evolution in a purely matter-dominated universe. In the DE models, the local expansion-rate is suppressed relative to EdS, with deviations appearing only at low redshift and disappearing at high redshift, where DE is negligible. Since we compare voids reaching the same depth at different epochs, the decreasing trend in DE models directly reflects the DE-induced suppression of structure growth. The parameter dependence follows the same interpretation discussed above. We stress that, at fixed redshift and void depth, EdS always gives a larger expansion-rate excess, even though the DE models require more negative ICs to reach the same final density contrast.

The main conclusion of this section is that the local Hubble excess generated by a void is primarily a matter-driven effect. Dynamical DE provides a secondary correction by changing the epoch at which the matter-driven growth of the void is suppressed. These results motivates treating DE dynamics as a correction to the leading matter-driven dependence of the local expansion rate.

\subsection{Implications for the Hubble tension}
\label{Sec:implications_for_the_Hubble_tension}

We now apply the hydrodynamical framework to the Hubble-tension problem. 
Specifically, we ask which present-day void depth is required for a Planck-calibrated 
background expansion to reproduce the local value of the Hubble constant inferred by 
SH0ES. We therefore fix the background normalization to 
\(H_0^{\rm P18}=67.4\,{\rm km\,s^{-1}\,Mpc^{-1}}\) and interpret the SH0ES value, 
\(H_0^{\rm SH0ES}=73.17\,{\rm km\,s^{-1}\,Mpc^{-1}}\), as the expansion rate measured 
inside a local underdensity at \(z=0\), i.e.
\begin{align}
    H_0^{\rm SH0ES} = H_{\rm v}(z=0)
    = \mathcal{H}_{\rm v}(z=0)\,H_0^{\rm P18}\,.
\end{align}
This matching condition must be interpreted as an idealized, leading-order estimate of the local expansion shift associated with a central observer embedded in a spherical underdensity, rather than as a forward model of the SH0ES distance-ladder analysis. Indeed, the cosmic distance ladder does not directly measure the instantaneous expansion rate at $z=0$, but rather infers an effective $H_0$ from the distance--redshift relation over a finite redshift interval (see, e.g.,~\cite{Redlich:2014gga,Efstathiou:2021ocp,Camarena:2022iae}). A fully consistent assessment would therefore require propagating a specific density and velocity profile through the supernova redshift distribution, selection function, and peculiar-velocity corrections, and reproducing the corresponding distance-ladder inference as in~\cite{Kenworthy:2019qwq}. However, across the range of $(w_0,w_a)$ values explored here, motivated by the observational results of DES and DESI, our leading-order estimate shows that fully reconciling the local and CMB-inferred values of $H_0$ would require a void that is approximately $50\%$ deeper than even the most extreme observational estimate of the local underdensity around us~\cite{Keenan:2013mfa}. We therefore do not expect a more detailed modeling of the distance-ladder observable to qualitatively alter our conclusions, but rather only to refine the precise magnitude of the residual tension. For this reason, we regard such an analysis as beyond the scope of the present work.

We first perform this exercise in a $\Lambda$CDM background, fixing 
$w_0=-1$ and $w_a=0$ and varying the present-day matter fraction
$\Omega_{\rm m,0}$. 
\begin{figure}[t]
    \centering
    \includegraphics[width=0.8\linewidth]{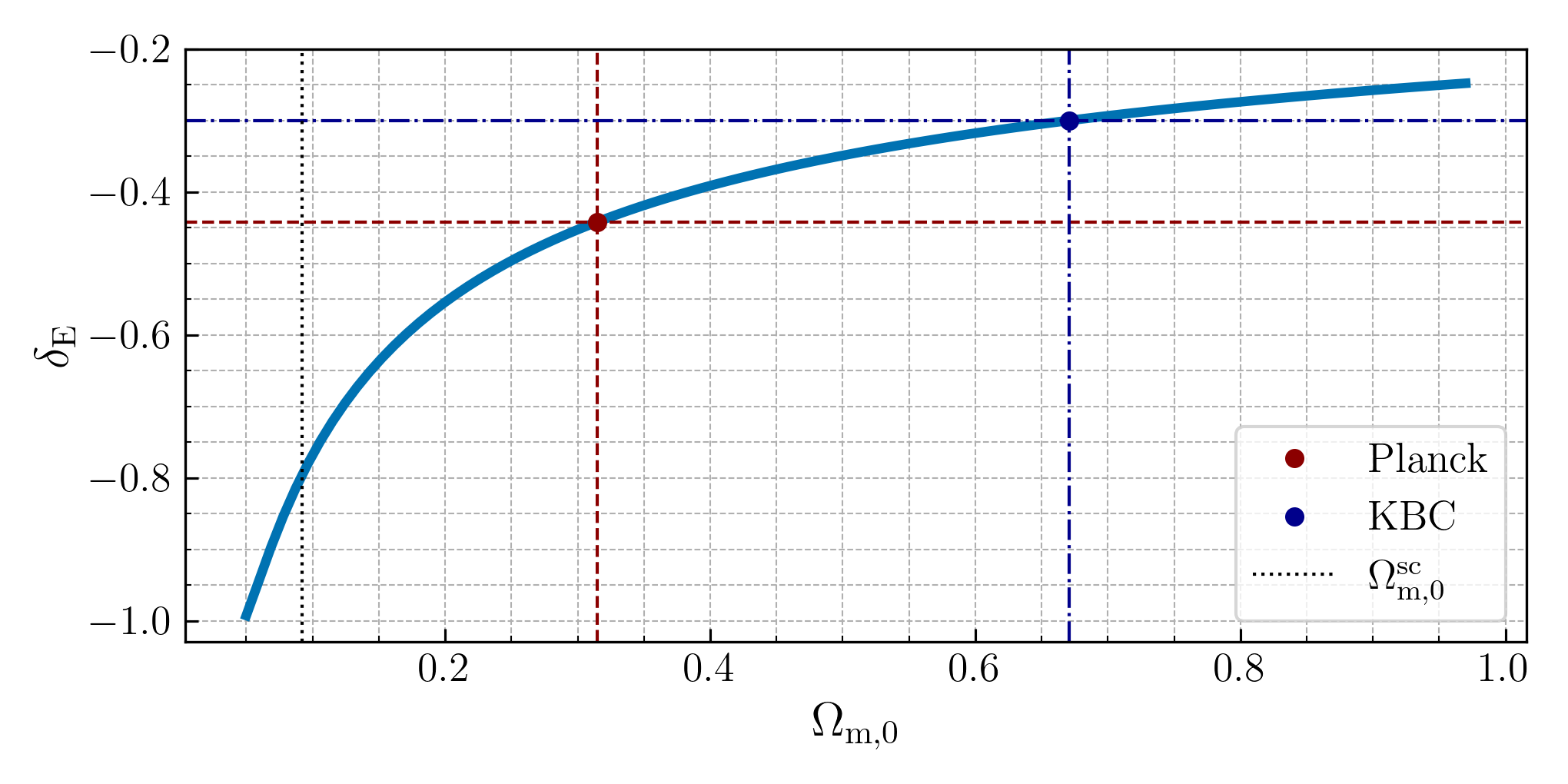}
    \caption{Required present-day void depth $\delta_{\rm E}(z=0)$ needed to reproduce the SH0ES value as a function of $\Omega_{\rm m,0}$ in a $\Lambda$CDM background. The background Hubble constant is fixed to the Planck best-fit value, $H_0^{\rm P18}=67.4\,{\rm km\,s^{-1}\,Mpc^{-1}}$. The red point marks the Planck-calibrated case, while the blue point marks the benchmark adopted for the KBC void, $\delta_{\rm E}^{\rm KBC}\sim -0.3$. The vertical black dotted line indicates $\Omega_{\rm m,0}^{\rm sc}$, where the required solution reaches shell-crossing at $z=0$.
}
    \label{fig:delta_H0_omega_0m}
\end{figure}
Figure~\ref{fig:delta_H0_omega_0m} shows the degeneracy between the present-day matter fraction and the void depth required to generate the fixed local expansion excess $H_0^{\rm SH0ES}/H_0^{\rm P18}-1 \simeq 8.6\%$. Each point on the curve therefore corresponds to a pair $(\Omega_{\rm m,0},\delta_{\rm E})$ yielding the same value of $H_{\rm v}(z=0)$. At fixed void depth, a larger matter fraction enhances the matter-driven evacuation of the underdensity and produces a larger local expansion-rate excess. Consequently, as $\Omega_{\rm m,0}$ increases, a progressively shallower void is sufficient to reproduce the SH0ES value.
For the Planck-preferred value $\Omega_{\rm m,0}=0.315$, the required depth is
$\delta_{\rm E}(z=0)\simeq -0.442$. In the following, we use
$\delta_{\rm E}^{\rm KBC}\equiv -0.3$ as an effective benchmark for the enclosed
matter contrast associated with a KBC-like local underdensity. With this
identification, the depth required to reproduce the SH0ES value is
significantly larger than the KBC-like benchmark.
Conversely, imposing a KBC-like depth would require $\Omega_{\rm m,0}\simeq 0.671$, far above the Planck-preferred value. The blue point should therefore not be interpreted as an observationally favoured cosmology, but rather as the intersection between the assumed KBC depth and the curve required to reproduce the SH0ES value.
The vertical dotted line marks the value of $\Omega_{\rm m,0}$ below which the required solution has already reached shell crossing by $z=0$. To the left of this limit, the single-stream hydrodynamical description adopted here is no longer self-consistent. Overall, the figure shows that, within a Planck-calibrated $\Lambda$CDM background, a KBC-like void is not deep enough to account for the full local expansion excess inferred by SH0ES.

\begin{figure}
    \centering
    \includegraphics[width=0.8\linewidth]{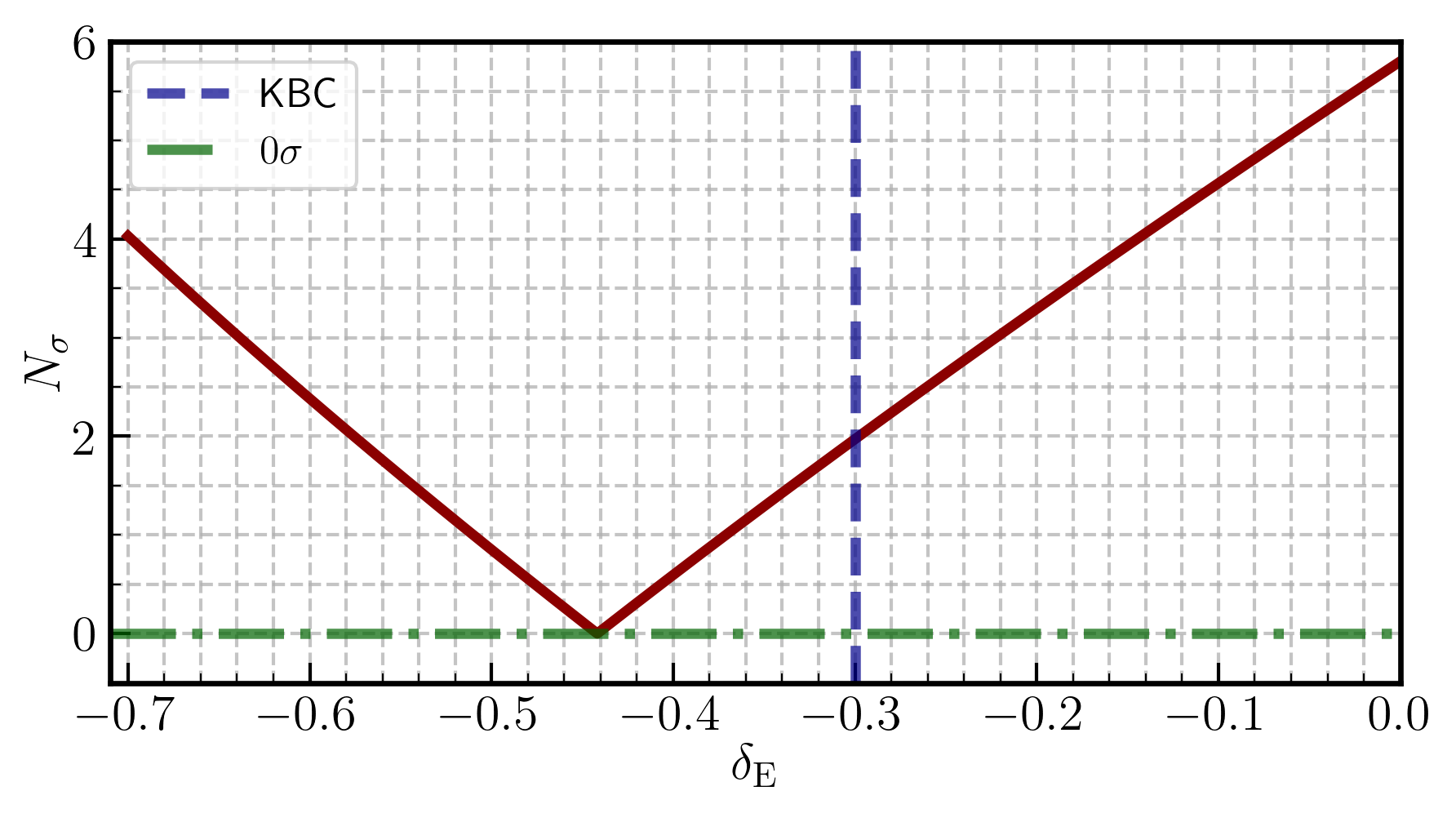}
    \caption{Residual tension with the CMB-calibrated value of $H_0^{\rm P18}$ as a function of the assumed present-day void depth $\delta_{\rm E}(z=0)$. 
    The dashed vertical line marks the KBC benchmark, $\delta_{\rm E}^{\rm KBC}\simeq -0.3$, while the horizontal line indicates zero residual tension.}
    \label{fig:sigma_delta}
\end{figure}

We now complement the previous analysis by fixing $\Omega_{\rm m,0}=0.315$ and quantifying the residual tension as a function of the assumed present-day void depth. Figure~\ref{fig:sigma_delta} shows the statistical discrepancy with the Planck-calibrated value of $H_0$ after correcting the SH0ES measurement for the void-induced local expansion excess, i.e.
\begin{align}
    H_0^{\rm corr}(\delta_{\rm E})
    =
    \frac{
    H_0^{\rm SH0ES}
    }{
    \mathcal{H}_{\rm v}(z=0,\delta_{\rm E})
    }\,.
\end{align}
Since $\mathcal{H}_{\rm v}>1$ inside an underdense region, this correction removes
the local expansion enhancement induced by the void and therefore lowers the
value of $H_0$ inferred from the local distance ladder. The residual discrepancy with the Planck-calibrated value is then quantified as
\begin{align}
    N_\sigma(\delta_{\rm E})
    =
    \frac{
    \left|H_0^{\rm corr}(\delta_{\rm E})-H_0^{\rm P18}\right|
    }{
    \sqrt{
    \left(\frac{\sigma_{\rm SH0ES}}{\mathcal{H}_{\rm v}(z=0,\delta_{\rm E})}\right)^2
    +
    \sigma_{\rm P18}^2
    }
    }\,,\qquad \sigma_{\rm P18}=0.5\,,\,\,\, \sigma_{\rm SH0ES}=0.86\,.
\end{align}
Here, the SH0ES uncertainty is propagated through the void correction,
while $\mathcal{H}_{\rm v}$ is treated as fixed for a given cosmological model
and void depth. For $\delta_{\rm E}=0$, no void correction is applied and the original
SH0ES--Planck tension is recovered. As the void becomes deeper, the local expansion enhancement increases and $H_0^{\rm corr}$ moves progressively closer to $H_0^{\rm P18}$, reducing the residual tension. The minimum occurs at $\delta_{\rm E}\simeq-0.44$, where $H_0^{\rm corr}\simeq H_0^{\rm P18}$ and the discrepancy therefore vanishes, consistently with the result shown in figure~\ref{fig:delta_H0_omega_0m}.
For voids deeper than $\delta_{\rm E}\simeq-0.44$, the correction becomes too large and yields $H_0^{\rm corr}<H_0^{\rm P18}$. Since $N_\sigma$ depends on the absolute difference between the two values, the residual tension increases again, explaining the shape of the curve around its minimum.
The vertical dashed line marks the KBC benchmark,
$\delta_{\rm E}^{\rm KBC}\simeq-0.3$. At this depth, the tension is reduced to approximately $2\sigma$, but it is not fully removed. More generally, the figure shows that the residual tension remains below $3\sigma$ over the range $\delta_{\rm E}\in[-0.64,-0.22]$. Therefore, a KBC-like void can alleviate the discrepancy between SH0ES and Planck, while a substantially deeper underdensity would be required to reconcile the two values completely within this framework.

\begin{figure}[t]
    \centering
    \includegraphics[width=0.8\linewidth]{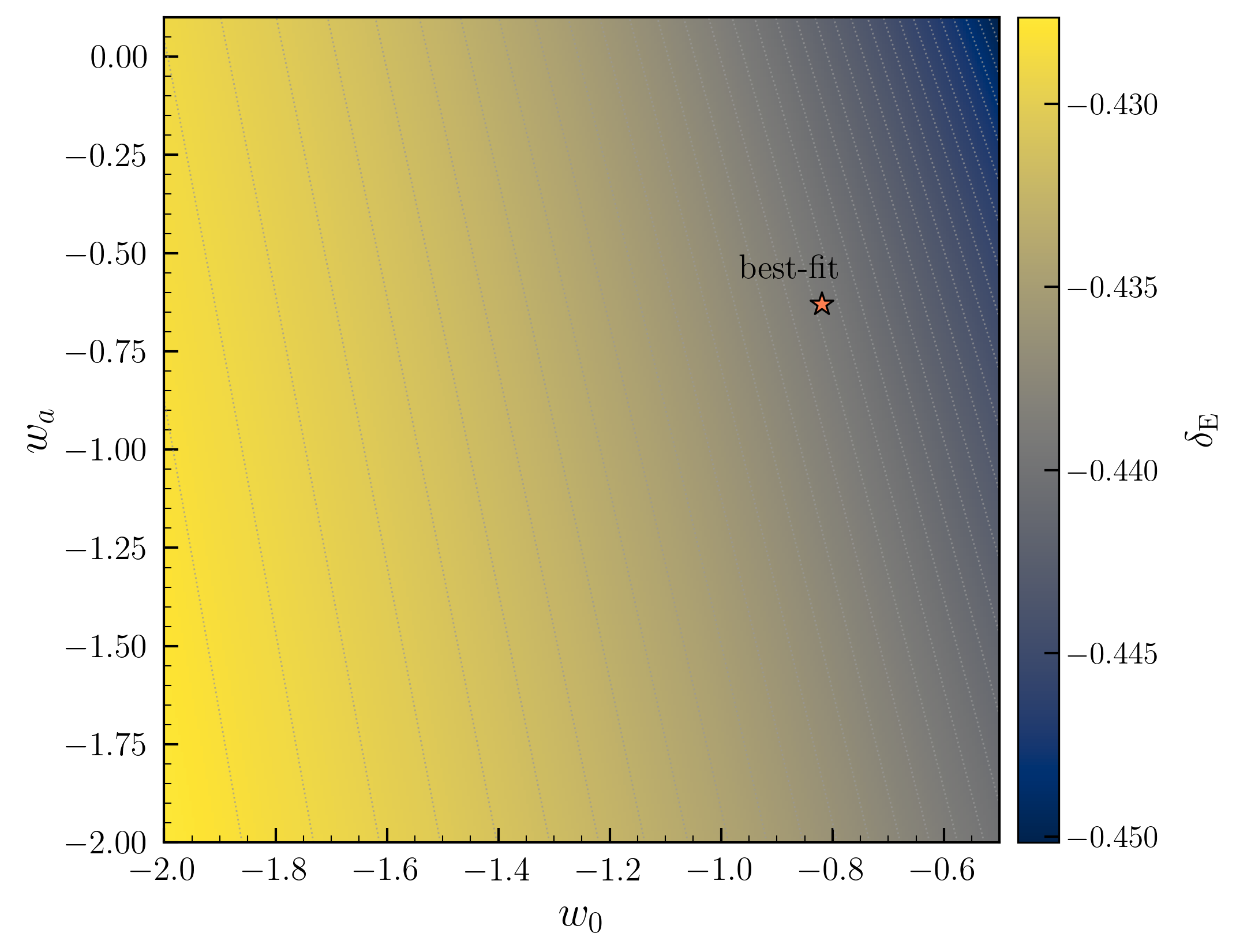}
    \caption{
    Required present-day void depth $\delta_{\rm E}(z=0)$ needed to reproduce the SH0ES value in dynamical DE backgrounds. 
    The matter fraction and background normalization are fixed to $\Omega_{\rm m,0}=0.32$ and $H_0=H_0^{\rm P18}$, while $w_0$ and $w_a$ are varied. The star marks the best-fit values of $(w_0,w_a)$ from the combined DESI+DES+CMB analysis reported in eq.~\eqref{eq:w0_wa_best_fit_DESI_DES_CMB}.
    }
    \label{fig:delta_H0_EOS_DE_01}
\end{figure}

We now repeat the matching procedure of figure~\ref{fig:delta_H0_omega_0m} for dynamical DE backgrounds, with the results shown in figure~\ref{fig:delta_H0_EOS_DE_01}.
We fix $\Omega_{\rm m,0}=0.32$ and the background Hubble normalization to 
$H_0^{\rm P18}$, while varying the DE parameters in the ranges 
$w_0\in[-2.0,-0.5]$ and $w_a\in[-2.0,+0.1]$.
The trend of $\delta_{\rm E}$ with $w_0$ and $w_a$ has the same physical origin discussed above: these parameters control the onset of DE domination and hence the late-time efficiency of void evacuation. Quantitatively, the required depth varies only at the percent level with respect to the corresponding $\Lambda$CDM value. This is consistent with the analysis of section~\ref{sec:de_effects}, where the effect of DE on $H_{\rm v}$ was found to be subdominant with respect to the dependence on the void depth and on $\Omega_{\rm m,0}$.
The conclusion is therefore unchanged. Over the range of $(w_0,w_a)$ considered here, the required underdensity remains substantially deeper than $\delta_{\rm E}^{\rm KBC}\simeq -0.3$. Dynamical DE can slightly shift the required value of $\delta_{\rm E}$, but it does not make a KBC-like local void sufficient to reproduce the SH0ES measurement in this setup.

We finally present two applications of the hydrodynamical model to recent DES+DESI constraints. 
In the first one, we restrict to late-time probes and use the DES+DESI results~\cite{DES:2026jmi}. 
We fix the matter fraction to the corresponding best-fit value, $\Omega_{\rm m,0}=0.31$, while we vary $w_0$ and $w_a$ within their reported $1\sigma$ intervals,
\begin{align}
    w_0 = -0.84^{+0.06}_{-0.07}\,,
    \qquad
    w_a = -0.53^{+0.33}_{-0.28}\,.
    \label{eq:w0_wa_best_fit_DES_DESI}
\end{align}
We then assume that our location coincides with the center of a KBC-like underdensity and ask which early-time normalization of the Hubble constant would be required for the void-induced local expansion rate to match the SH0ES measurement, i.e.
\begin{align}
    H_0^{\rm early}(w_0,w_a)
    =
    \frac{H_0^{\rm SH0ES}}
    {\mathcal{H}_{\rm v}(\delta_{\rm E}^{\rm KBC},w_0,w_a)}\, .
\end{align}
Second, we consider the combined DES+DESI+CMB constraints. We adopt the best-fit values for $\Omega_{\rm m,0}$ and $H_0$, i.e. $\Omega_{\rm m,0}=0.312$ and $H_0=67.4\,{\rm km\,s^{-1}\,Mpc^{-1}}$, and vary the DE EoS parameters within their quoted $1\sigma$ intervals,
\begin{align}
    w_0=-0.82\pm0.05\,,
    \qquad
    w_a=-0.63^{+0.21}_{-0.18}\,.
    \label{eq:w0_wa_best_fit_DESI_DES_CMB}
\end{align}
For each point in this $(w_0,w_a)$ region, we then ask which present-day void depth, $\delta_{\rm E}$, is required for the void-induced local expansion rate to reproduce the SH0ES measurement.

\begin{figure}[t]
    \centering
    \includegraphics[width=1.0\linewidth]{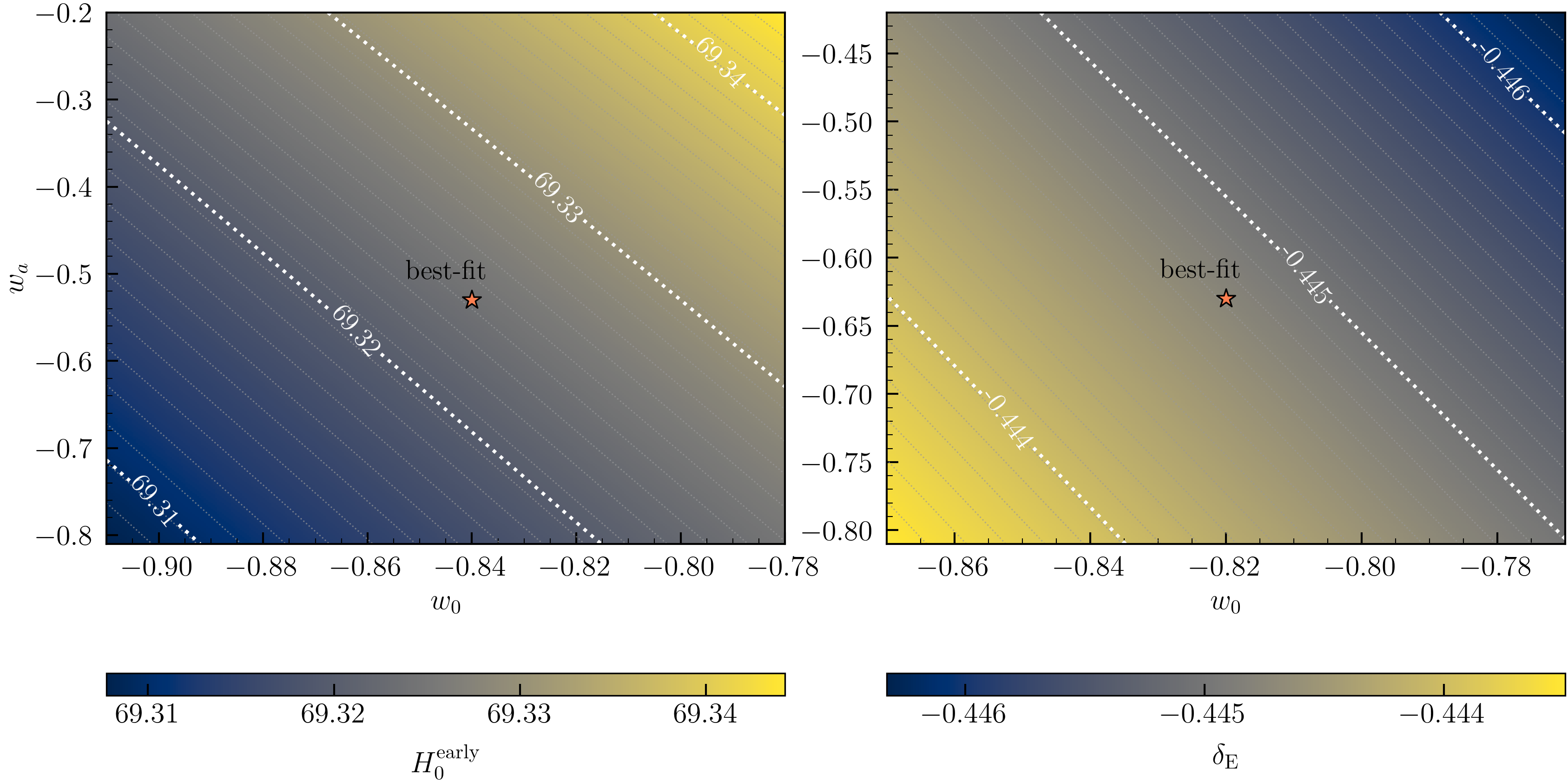}
    \caption{Left panel: $H_0^{\rm early}$ required for a KBC-like underdensity to reproduce $H_0^{\rm SH0ES}$, with $\Omega_{\rm m,0}=0.31$ fixed to the DES+DESI best-fit value and $(w_0,w_a)$ varied within the corresponding $1\sigma$ region. 
    Right panel: $\delta_{\rm E}$ at $z=0$ required to reproduce $H_0^{\rm SH0ES}$, with $\Omega_{\rm m,0}=0.312$ and $H_0=67.4\,{\rm km\,s^{-1}\,Mpc^{-1}}$ fixed to the best-fit values of the DES+DESI+CMB combination and $(w_0,w_a)$ varied within the corresponding $1\sigma$ region. The star marks the best-fit $(w_0,w_a)$ values inferred from the DESI+DES combination, eq.~\eqref{eq:w0_wa_best_fit_DES_DESI}, in the left panel, and DESI+DES+CMB combination, eq.~\eqref{eq:w0_wa_best_fit_DESI_DES_CMB}, in the right panel.}
    \label{fig:double_analysis}
\end{figure}
The results are shown in figure~\ref{fig:double_analysis}, application 1 (left) and application 2 (right) respectively. In both cases, the dependence on $(w_0,w_a)$ follows the same interpretation discussed previously: the DE EoS controls the onset of DE domination and therefore the late-time efficiency of void evacuation. However, the induced variations remain below the percent level. The dominant effect is therefore controlled by the matter sector and by the void depth, while dynamical DE gives only a subdominant correction.
In the left panel, a KBC-like underdensity requires 
$H_0^{\rm early}\simeq 69.31-69.34\,{\rm km\,s^{-1}\,Mpc^{-1}}$, 
above the CMB-calibrated normalization. 
Thus, even allowing for dynamical DE within the DES+DESI region, such a void can alleviate the Hubble tension but cannot remove it. 
The right panel gives the complementary result: within the DES+DESI+CMB-favored CPL region, the depth required to account for $H_0^{\rm SH0ES}$ remains well below the KBC benchmark.


\subsection{Redshift dependence of the locally inferred Hubble rate}
\label{Sec:redsfhit_dependence_of_the_locally_inferred_Hubble_rate}

Several analyses have investigated whether the Hubble constant inferred from late-time probes may exhibit an effective redshift dependence~\cite{Krishnan:2020obg,Krishnan:2020vaf,Dainotti:2021pqg,Jia:2022ycc,Jia:2024wix}. In particular, building on ref.~\cite{Jia:2022ycc}, the analysis of ref.~\cite{Jia:2024wix} constrained the evolution of $H_0$ in a flat $\Lambda$CDM background by combining cosmic-chronometer measurements of $H(z)$ from~\cite{Moresco:2023zys}, DESI BAO measurements~\cite{DESI:2024mwx}, and the Pantheon+ SNe Ia sample~\cite{Brout:2022vxf}. They reported evidence for a decreasing trend of $H_0$ with redshift at the level of $6.4\sigma$.
This inference is based on a non-parametric description of the redshift dependence of $H_0$. Instead of assuming a specific functional form, $H_0(z)$ is taken to be constant within redshift bins,
\begin{align}
    H_0(z) =
    \begin{cases}
        H_{0,z_1}\,, & \text{if } 0 \le z < z_1\,, \\
        \cdots & \cdots\\
        H_{0,z_i}\,, & \text{if } z_{i-1} \le z < z_i\,,\\
        \cdots & \cdots\\
        H_{0,z_N}\,, & \text{if } z_{N-1} \le z \le z_N \,,
    \end{cases}
    \label{eq:H_0_z_i}
\end{align}
where $N$ is the total number of redshift bins and $H_{0,z_i}$ is the value of $H_0(z)$ in the $i$-th bin. The expansion rate is therefore written as
\begin{align}
    H(z)
    =
    H_{0}(z)\,E(z)= H_{0}(z)\sqrt{\Omega_{\rm m,0}(1+z)^3 + \Omega_{\Lambda,0}}\,,
    \label{eq:Hz_H0_zi}
\end{align}
where $E(z)$ is the dimensionless expansion rate and $\Omega_{\Lambda,0}$ is the present-day density parameter associated with $\Lambda$.
The analysis assumes a flat background with no radiation. Since the data do not provide sufficient constraining power to fit $\Omega_{\rm m,0}$ simultaneously with the binned values of $H_0$, this parameter is fixed to $\Omega_{\rm m,0}=0.315$. The robustness of the decreasing trend was tested against changes in both $\Omega_{\rm m,0}$ and the redshift binning, with the statistical significance depending on these choices.
In what follows, we use the results of ref.~\cite{Jia:2024wix} reported in table~\ref{tab:H_0_z_i}, which lists the redshift bins adopted in their non-parametric inference together with the corresponding best-fit values of $H_{0,z_i}$ and their $1\sigma$ uncertainties.
\begin{table}[t!]
    \centering
    \caption{Fitting results of $H_{0,z_i}$ used in this work, taken from table I of~\cite{Jia:2024wix}. 
    The values are given in units of $\mathrm{km\,s^{-1}\,Mpc^{-1}}$.}
    \label{tab:H_0_z_i}
    \begin{tabular}{ccccc}
        \toprule
        \toprule
        Redshift bin & $H_{0,z_i}$ \\
        \midrule
        $[0,0.10]$    & $73.28_{-0.14}^{+0.14}$ \\
        \addlinespace[2pt]
        $[0.10,0.20]$ & $72.96_{-0.31}^{+0.31}$ \\
        \addlinespace[2pt]
        $[0.20,0.30]$ & $72.39_{-0.46}^{+0.47}$ \\
        \addlinespace[2pt]
        $[0.30,0.40]$ & $70.49_{-0.66}^{+0.66}$ \\
        \addlinespace[2pt]
        $[0.40,0.60]$ & $70.52_{-0.63}^{+0.67}$ \\
        \addlinespace[2pt]
        $[0.60,0.80]$ & $67.33_{-0.99}^{+1.00}$ \\
        \addlinespace[2pt]
        $[0.80,1.00]$ & $64.85_{-1.00}^{+1.01}$ \\
        \addlinespace[2pt]
        $[1.00,1.50]$ & $68.14_{-1.60}^{+1.64}$ \\
        \bottomrule
        \bottomrule
    \end{tabular}
\end{table}

We stress that the quantities $H_{0,z_i}$ are effective parameters obtained by
fitting the distance--redshift relation independently in different redshift bins.
They should not be interpreted as direct measurements of a time-dependent
background value of $H_0$. In the following, we investigate whether their
redshift dependence can instead be represented phenomenologically as a
modulation of the local expansion induced by the enclosed matter distribution.

Physically, this amounts to interpreting the inferred redshift dependence of $H_0$ as an effective modulation of the locally measured expansion rate induced by the surrounding matter field. In this interpretation, the binned quantities $H_{0,z_i}$ are not taken to represent values of the background value of $H_0$ at different redshifts. Rather, each $H_{0,z_i}$ is used to infer the enclosed matter contrast required to reproduce the corresponding effective expansion rate within the sphere whose radius is set by that redshift bin. We then compare the reconstructed matter profile against observational probes of the local matter field.

Operationally, we perform a tomographic reconstruction of the local matter field profile using the hydrodynamical model introduced in section~\ref{sec:the_model}. We first fix the background cosmology to the Planck-calibrated flat $\Lambda$CDM model used above, with $\Omega_{\rm m,0}=0.315$, no radiation, and $H_0=67.4\,{\rm km\,s^{-1}\,Mpc^{-1}}$. 
Then, for each of the eight redshift bins, we introduce a spherical top-hat shell centered on the observer. We place each shell at the central redshift of the corresponding bin,
\begin{align}
    \bar z_i
    =
    \frac{z_{i-1}+z_i}{2}\,,
    \label{eq:central_redshift}
\end{align}
with $i=1,\ldots,8$. This assignment should be understood as an effective simplifying prescription. Since each $H_{0,z_i}$ is inferred from data distributed over a finite redshift interval, assigning it to a single shell requires choosing a representative redshift. Assuming that the effective expansion normalization does not vary sharply within each bin, the bin midpoint provides a natural approximation to the characteristic scale probed by that bin. We then identify the physical radius of the shell at $\bar z_i$ with the physical distance between the observer and that redshift,
\begin{align}
    R(\bar z_i)
    = \frac{d_{\rm comov}(\bar z_i)}{1+\bar z_i}\,,
    \label{eq:shell_radius_matching}
\end{align}
where $R$ is the physical radius of the shell and $d_{\rm comov}$ is the comoving distance between the observer and the redshift $\bar z_i$.~\footnote{We compute comoving distances using \texttt{Astropy}~\cite{astropy:2013,astropy:2018,astropy:2022}.} 
Thus, for each shell, we determine the value of $\Delta_{\rm E}(\bar z_i)$ required to reproduce the corresponding binned expansion rate. We repeat the procedure for the central value of $H_{0,z_i}$ and for its upper and lower $1\sigma$ limits. This is done by shooting on the ICs, i.e. $\delta_{\rm v,in}$ in eq.~\eqref{Eq:initial_condition_delta_prime}, until
\begin{align}
    \mathcal{H}_{\rm loc}(\bar z_i, \Delta_{\rm E}(\bar z_i))\,H_0
    =
    H_{0,z_i}\,,
    \label{eq:H0zi_target}
\end{align}
where we recall that $\mathcal{H}_{\rm loc}$ is the local expansion rate normalized to the background one defined in eq.~\eqref{eq:Hv_def}.
The condition in eq.~\eqref{eq:H0zi_target} can be imposed because the same background factor $E(z)$ appears in both the binned expansion rate of eq.~\eqref{eq:Hz_H0_zi} and the local one of eq.~\eqref{eq:Hv_def}. 

Before presenting the results, we stress three points about the reconstruction. First, the sign of the ICs is fixed by comparing each $H_{0,z_i}$ with the
background value $H_0$. For bins with $H_{0,z_i}>H_0$, the matching requires
negative ICs and therefore a negative enclosed density contrast, corresponding
to an underdense region. Conversely, bins with $H_{0,z_i}<H_0$ formally require
positive ICs and a positive enclosed density contrast. The hydrodynamical
treatment is unchanged in this case (see, e.g.~\cite{Abramo:2007iu}): the same evolution
equation and shooting procedure apply, but the ICs are positive.
Such positive values may correspond to a compensating overdense environment
surrounding the inner underdensity. However, this interpretation is physically
meaningful only if all the inferred values of $\Delta_{\rm E}(<R)$ can be embedded
consistently within the same continuous cumulative mass profile. A change of
sign in $\Delta_{\rm E}(<R)$ should therefore not, by itself, be interpreted as
evidence for a compensating shell.

Second, the matching is performed independently for each redshift bin, whereas the corresponding spherical shells are nested and must ultimately belong to a single matter distribution. The inferred values of the enclosed contrast $\Delta_{\rm E}(<R_i)$ are therefore not physically independent. Finding a solution for each bin separately does not guarantee the existence of a continuous density profile $\delta_{\rm E}(R)$ that reproduces all bins simultaneously. The reconstruction should consequently be regarded as a consistency test: after the matching, the enclosed contrasts must be checked for mutual compatibility and for the absence of unphysical mass redistribution or excessively sharp variations between neighbouring shells.

Third, the matching condition in eq.~\eqref{eq:H0zi_target} is independent of the absolute shell radius. As discussed in section~\ref{sec:the_model}, the hydrodynamical model fixes the density evolution but is insensitive to the absolute radius normalization, which must be specified separately. The radius assigned through eq.~\eqref{eq:shell_radius_matching} is therefore not an input of the shooting procedure, but a prescription to attach a physical scale to the reconstructed values of $\Delta_{\rm E}$. This scale information is needed to assess whether the resulting enclosed matter profile can be physically realized and compared with observations.
\begin{figure}[t]
    \centering
    \includegraphics[width=1.0\linewidth]{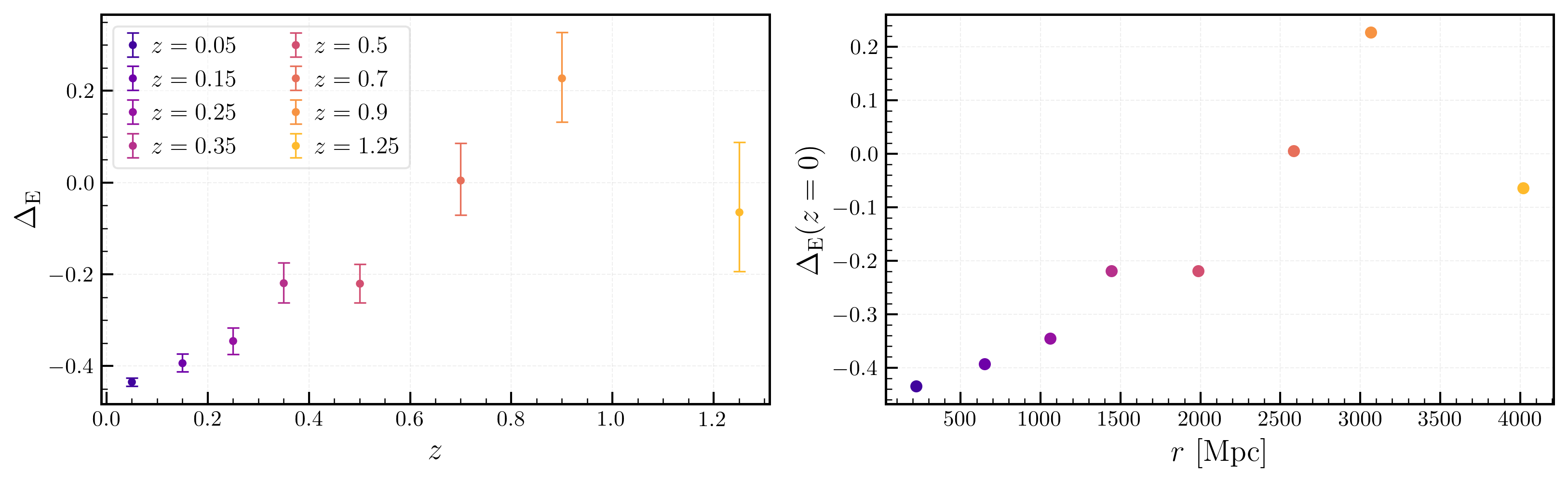}
    \caption{Tomographic reconstruction of the local matter field profile required to reproduce the binned values of $H_{0,z_i}$ reported in table~\ref{tab:H_0_z_i}. 
    Left panel: enclosed density contrast $\Delta_{\rm E}(\bar z_i)$ inferred at the central redshift of each bin. Error bars are obtained by propagating the upper and lower $1\sigma$ uncertainties on $H_{0,z_i}$. 
    Right panel: corresponding enclosed matter profile evolved to $z=0$. For each shell, the solution that matches the central value of $H_{0,z_i}$ at $\bar z_i$ is evolved to the present time using the hydrodynamical model, and the final enclosed density contrast $\Delta_{\rm E}(z=0)$ is shown as a function of the final physical radius $R(z=0)$. 
    Colors label the central redshift of each bin.}\label{fig:void_profile_hubble_data}
\end{figure}
The reconstructed quantity is the averaged density contrast $\Delta_{\rm E}(<R)$, rather than the local density contrast $\delta_{\rm E}(R)$. Recovering the latter would require differentiating the enclosed mass profile and would amplify the
bin-to-bin noise. We therefore restrict the interpretation to the cumulative profile.

We present our results in figure~\ref{fig:void_profile_hubble_data}. The left panel shows the enclosed density contrast $\Delta_{\rm E}(\bar z_i)$ required to reproduce the effective expansion rate inferred in each redshift bin. Each point corresponds to the central redshift $\bar z_i$ of one bin, while the error bars are obtained by propagating the upper and lower $1\sigma$ uncertainties on $H_{0,z_i}$ through the shooting procedure. The colors label the redshift bins. The right panel shows the corresponding matter profile at $z=0$. For each shell, we evolve $R(\bar z_i)$ and $\Delta_{\rm E}(\bar z_i)$ from $\bar z_i$ to the present time using the hydrodynamical model, and plot the reconstructed $\Delta_{\rm E}$ as a function of the resulting present-day radius. For clarity, we show only the profile obtained from the central values of $H_{0,z_i}$. Propagating the $1\sigma$ uncertainties in this panel would shift each point both vertically, through $\Delta_{\rm E}$, and horizontally, through the corresponding radius evolution, making the radial trend less transparent.

The purpose of this reconstruction is not to claim that it provides a solution to the phenomenological inputs of~\cite{Jia:2024wix}. Rather, it is to show that the reported redshift dependence of the locally inferred $H_0$ across different redshift bins can be always formally reproduced as an effective modulation induced by the enclosed matter distribution. Figure~\ref{fig:void_profile_hubble_data} shows that this matching can be achieved with an enclosed density profile that does not display extreme bin-to-bin variations. In this sense, the reconstructed profile is phenomenologically sensible. A quantitative assessment of its viability, however, requires a direct comparison with independent measurements of the local matter distribution on the same scales, which are currently not precise enough to provide a stringent bin-by-bin test.

We note that the innermost bin requires an enclosed underdensity close to the value found in section~\ref{Sec:implications_for_the_Hubble_tension}, namely $\delta_{\rm E}\simeq -0.44$. This is expected, since the lowest-redshift bin is strongly tied to the local SH0ES measurements. Therefore, although the reconstructed matter profile can formally reproduce the decreasing trend of $H_{0,z_i}$ without extreme bin-to-bin variations, the depth required in the innermost region is already disfavored by the comparison with KBC-like underdensities discussed above. Within the present setup, this makes a purely matter-profile interpretation of the full trend difficult to support.

However, this analysis highlights that the interpretation of the binned quantities $H_{0,z_i}$ is itself non-trivial. If the matter distribution between the observer and the sources in a given bin is not described by the homogeneous background, part of its effect can be absorbed into the effective value of $H_{0,z_i}$ inferred from the distance-redshift relation. Therefore, the observed trend should not be immediately interpreted as a genuine redshift dependence of the background value of $H_0$ without first disentangling it from local-structure effects.

\section{Conclusions}
\label{sec:conclusions}

In this work, we studied the local expansion rate inside isolated spherical cosmic voids using the hydrodynamical model for matter perturbations~\cite{Moretti:2025gbp}, based on the continuity, Euler, and Poisson equations. We applied the model to inverse top-hat underdensities evolving in a homogeneous and isotropic background. Within this framework, the void dynamics is fully specified by the evolution of the enclosed density contrast, while the local expansion rate follows from mass conservation. This provides a direct way to compute the void Hubble shift, $\Delta H_{\rm v}$, namely the fractional excess of the local Hubble rate with respect to the background. 

We applied the framework to flat $w_0w_a$CDM cosmologies, using EdS and $\Lambda$CDM as reference limits. This allowed us to quantify the void Hubble shift as a function of redshift, void depth, and cosmological parameters, $\Delta H_{\rm v}(\delta_{\rm E},z,\Omega_{\rm m,0},w_0,w_a)$. For a single void trajectory, i.e. $\delta_{\rm E}=\delta_{\rm E}(z)$, we find that DE suppresses the matter-driven evacuation of the void at low redshift, and $\Delta H_{\rm v}$ develops a turnover, in contrast with the monotonic EdS behavior.
We then studied $\Delta H_{\rm v}$ both at fixed redshift and at fixed void depth. The most relevant case is $\Delta H_{\rm v}(z=0)$, for which we find that the dominant dependence is controlled by the matter sector, through the void depth $\delta_{\rm E}$ and the background matter fraction $\Omega_{\rm m,0}$. The CPL parameters provide secondary corrections: $w_0$ gives a milder modulation, while $w_a$ remains subdominant over the range explored and depends on the chosen value of $w_0$. Relative to $\Lambda$CDM, the CPL correction acts mainly as an overall rescaling of $\Delta H_{\rm v}(z=0)$ rather than as a change in its dependence on void depth.

We next applied this machinery to address the Hubble tension problem in different background cosmologies. Across these backgrounds, we asked which present-day void depth is required to reproduce the SH0ES value through the void-induced local Hubble shift. In $\Lambda$CDM cosmologies, fixing $H_0^{\rm P18}=67.4\,{\rm km\,s^{-1}\,Mpc^{-1}}$ and $\Omega_{\rm m,0}=0.315$ gives $\delta_{\rm E}(z=0)\simeq -0.442$. This is roughly $50\%$ deeper than the density contrast observationally claimed for the KBC-like underdensity around our position, $\delta_{\rm E}^{\rm KBC}\simeq -0.3$. Conversely, imposing a KBC-like depth would require $\Omega_{\rm m,0}\simeq 0.671$, well above the Planck-preferred value.
We also quantified the residual tension as a function of the assumed void depth. A KBC-like underdensity lowers the discrepancy to the $\sim 2\sigma$ level but does not remove it. More generally, the tension remains below $3\sigma$ for void depths in the approximate range $\delta_{\rm E}\in[-0.64,-0.22]$.

Allowing for dynamical DE changes these results only at the percent level. Varying $(w_0,w_a)$ over broad CPL ranges, as well as within regions motivated by recent DES and DESI analyses, shifts the required void depth but does not alter the main conclusion. 

Finally, we used the hydrodynamical model to investigate whether the reported redshift dependence of the effective binned values of $H_0$~\cite{Jia:2022ycc,Jia:2024wix} can be
represented phenomenologically as a modulation induced by the enclosed matter distribution. For each redshift bin, we inferred the enclosed density contrast required to reproduce the corresponding effective expansion rate. The resulting bin-by-bin matching does not, however, guarantee that all the inferred values can be embedded within a single continuous and physically admissible matter profile.
The reconstruction should therefore be regarded as a consistency test and must be confronted with independent observations of the local matter distribution, especially on the large scales associated with the highest-redshift bins. Our results suggest that this interpretation of the redshift evolution of binned $H_0$ values is difficult to support.

Overall, our analysis shows that the hydrodynamical model provides a simple and predictive framework to compute the local Hubble rate as a function of redshift, void depth, and cosmological parameters. Local underdensities can contribute to alleviating the Hubble tension and should be modelled carefully, together with survey selection and
peculiar-velocity corrections, in low-redshift determinations of $H_0$. However, within the spherical inverse top-hat setup considered here, KBC-like voids are not sufficient to resolve the tension.

\acknowledgments
We are deeply grateful to Giovanni for the many insightful discussions, valuable ideas, and generous exchanges that helped shape previous papers, as well as the present work.

T.M. acknowledges financial support from the Italian Space Agency (ASI) through the ASI-CAIF fellowship.
T.M. and N.F. acknowledge  the COST Action CosmoVerse, CA21136, supported by COST (European Cooperation in Science and Technology). T.M. and N.F. acknowledge the Istituto Nazionale di Fisica Nucleare (INFN) Sez. di Napoli, Iniziativa Specifica InDark.
T.M. and G.V. acknowledge support from the Simons Foundation to the Center for Computational Astrophysics at the Flatiron Institute. T.M. acknowledges the COST Action COSMIC WISPers, CA21106, and the COST Action BridgeQG, CA23130.

\bibliographystyle{apsrev4-2}
\bibliography{main}

\end{document}